\documentclass[11pt,draftcls,peerreviewca,onecolumn,a4paper,dvips]{IEEEtran}
\usepackage{tipa}
\usepackage{amsfonts}
\usepackage{amsfonts}
\usepackage{amsfonts}
\usepackage{mathbbold}
\usepackage{amsfonts}
\usepackage{amssymb}
\usepackage{stfloats}
\usepackage{cite}
\usepackage{graphicx}
\usepackage{psfrag}
\usepackage{subfigure}
\usepackage{amsmath}
\usepackage{array}

\DeclareMathOperator*{\argmin}{\arg\min}
\DeclareMathOperator*{\argmax}{\arg\max}
\DeclareMathOperator*{\ThetaL}{\mathbf{\Theta}}

\begin{document}

\title{
Partial Interference Alignment for $K$-user MIMO Interference Channels}

\newtheorem{Thm}{Theorem}
\newtheorem{Lem}{Lemma}
\newtheorem{Cor}{Corollary}
\newtheorem{Def}{Definition}
\newtheorem{Exam}{Example}
\newtheorem{Alg}{Algorithm}
\newtheorem{Prob}{Problem}
\newtheorem{Rem}{Remark}
\newtheorem{Proof}{Proof}
\newtheorem{Subproblem}{Subproblem}
\newtheorem{assumption}{Assumption}
\newtheorem{Subalgorithm}{Subalgorithm}

\author{\authorblockN{Huang Huang, {\em Student Member, IEEE}, Vincent K. N. Lau, {\em Senior Member, IEEE}}
\thanks{The authors are with the Department of Electronic and Computer Engineering
(ECE), Hong Kong University of Science and Technology (HKUST), Hong
Kong.}}

\maketitle

\begin{abstract}
In this paper, we consider a { Partial Interference Alignment and Interference Detection (PIAID) design} for $K$-user quasi-static MIMO interference channels with discrete constellation inputs. Each
transmitter has $M$ antennas and transmits $L$ independent data
streams to the desired receiver with $N$ receive antennas. { We focus on the case where not all $K-1$ interfering transmitters can be aligned at every receiver. As a result, there will be residual interference at each receiver that cannot be aligned. Each receiver detects and cancels the residual interference based on the constellation map.
However, there is a window of {\em unfavorable interference profile} at the receiver for Interference Detection (ID). In this paper, we propose a low complexity Partial Interference Alignment
scheme in which we dynamically select the user set for IA so as to create a {\em favorable interference profile} for ID at each receiver.} We first derive the
average { symbol error rate (SER)} by taking into account of the non-Guassian residual
interference { due to discrete constellation.} Using graph theory, we then devise a low complexity user set
selection algorithm for the PIAID scheme, { which minimizes the asymptotically tight bound for the average
end-to-end SER performance.} Moreover, we substantially simplify interference
detection at the receiver using Semi-Definite Relaxation (SDR)
techniques. It is shown that the SER performance of the proposed
PIAID scheme has significant gain compared with various conventional
baseline solutions.

\end{abstract}


\newpage
\section{Introduction}
{ Interference has been a very difficult problem in wireless communications.}. For instance, the capacity region of two-user
Gaussian interference channels has been an open problem for over 30
years \cite{HK:gaussian_IC}. Recently, there are some progress made
in understanding the interference
\cite{IC:gaussian:capacity,Maddah-Ali:2008} and extensive
studies have been done regarding the { {\em Interference Alignment} (IA) \cite{Maddah-Ali:2008}.}
For instance, IA is a signal processing approach that attempts to simultaneously align
the interference on a lower dimension subspace at each receiver so
that the desired signals can be transmitted on the {\em
interference-free} dimensions. In \cite{IA:conventional:2008}, the authors show that IA (using infinite dimension symbol extension in time or frequency selective fading channels) is optimal in Degrees-of-Freedom (DoF) sense. In
\cite{Ergodic:alignment:2009} the authors propose a variation of the
IA scheme, the ergodic alignment scheme, for $K$-user time or
frequency-selective interference channels. { In practice, since it is
not possible to realize infinite dimension symbol extensions, there
are a number of works \cite{IA:distributed:2008,maximum:sum:rate,alter:VT:2011} and the reference therein
that consider IA in the spatial domain, without symbol extensions,
in the $K$-user quasi-static MIMO interference channels. Furthermore, the authors in \cite{IA:analysis} investigate the asymptotic performance of these different IA solutions. However,
for quasi-static (or constant) channels, conventional IA might be infeasible
depending on the system parameters. For example, it is { conjectured} in
\cite{Feasibility:MIMO:2009}} that conventional IA on quasi-static
MIMO ($M$ transmit and $N$ receive antennas) interference channels
is not feasible to achieve a per user DoF greater than
$\frac{M+N}{K+1}$. As a result, we cannot rely on IA to eliminate
all interference in quasi-static MIMO interference channels
especially when $K$ is large. { There are other works that consider IA over signal scale in \cite{Many2one:2009,cadambe:lattice:2009,Real:2009,Real:MIMO:2009}.
In \cite{Many2one:2009} and \cite{cadambe:lattice:2009}, signal scale alignment schemes are introduced for the
many-to-one interference channel and fully connected interference networks, respectively. In \cite{Real:2009,Real:MIMO:2009}, the
authors propose a lattice alignment scheme for $K$-user MIMO
interference channels. However, the scheme requires infinite SNR and
serves only as a { proof of concept}. It is not clear whether this
approach can be applied at finite SNR.

Due to the fact that not all the interferers can be aligned at each receiver, there will be residual interference at the receiver. In this paper, we assume the receiver has interference detection (ID) capability. Specifically, the receiver detects and cancels the residual interference based on the constellation map  derived from the discrete constellation inputs. However, there is a window of {\em unfavorable interference profile} for ID at the receiver. For instance, ID at the receiver is more effective
when the interference is stronger than the desired signal \cite{Symmetric:2008,Real:MIMO:2009}.}
In \cite{Symmetric:2008,Threeusers:2008}, the authors propose an ID
scheme for quasi-static interference channels based on lattices. However the proposed scheme can only work under
idealized assumptions such as the symmetric SISO interference
channels (where all cross links have the same fading coefficients)
or a specialized class of 3-user SISO interference channels (where
the product of fading coefficients is assumed to be rational).

In fact, IA and ID are complementary approaches to deal with
interference in quasi-static MIMO interference channels. The IA
approach\footnote{ In the remaining papers, the mentioned IA approach specifically refers to the signal space alignment approach.} can be used to first eliminate some interference and the ID
approach can be used to deal with the residual interference at each
receiver. While we can potentially benefit from the concepts of IA
and ID approaches in dealing with interference, there are still some
key technical challenges to be addressed.
\begin{itemize}
\item{\bf Feasibility Issue of IA and Path Loss Effects:} { Sometimes brute-force IA in quasi-static MIMO interference channels (without symbol extensions) might be infeasible depending on system parameters}. Furthermore, existing literature has completely ignored the effects of path loss, which may also be exploited when dealing with interference. For instance, nodes with large path loss may not need to be interference-aligned and hence, it is important to jointly consider the feasibility issue and the path loss effects.

\item{\bf Coupling between IA and ID:} While both IA and ID are
effective means to mitigate interference, their designs are coupled
together in { an} intricate manner. For instance, the
performance of ID at the receiver depends heavily on the
interference profile. Hence,
IA can potentially contribute to creating a more desirable
interference profile for ID by careful selecting a subset of users
for IA. However, the problem of user selection in IA to optimize the
{ symbol error rate (SER) performance} is very complicated. First, the
optimization space is combinatorial and brute force exhaustive
search is not viable. Second, even if we can afford the search
complexity, obtaining the search metric is highly non-trivial
because it is also very challenging to analyze the closed form
average SER under non-Gaussian interferences.
\end{itemize}

\begin{figure}
\centering
\includegraphics[width = 16cm]{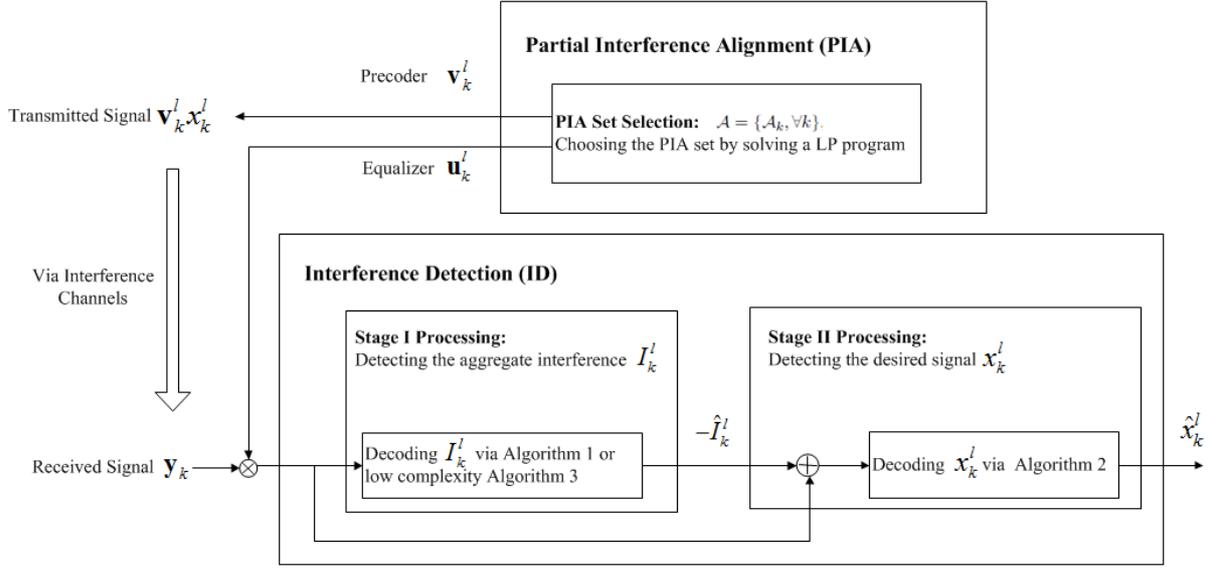}
\caption{ System level block diagram of the proposed PIAID Scheme.} \label{fig:PIAID_scheme}
\end{figure}

In this paper, we propose a low complexity {\em Partial Interference
Alignment} and {\em Interference Detection} (PIAID) scheme for
$K$-user quasi-static MIMO interference channels with discrete
constellation inputs. We consider QPSK constellations\footnote{ QPSK constellations are easier to analyze. However, the
proposed framework can be extended to QAM constellations easily.} at
the inputs of the $K$ transmitters and each transmit-receive pair
may have different path losses. The proposed PIAID scheme
dynamically selects the {\em interference alignment set} at each
receiver based on the path loss information { to create a {\em favorable interference profile} at each receiver for ID processing as illustrated in Fig. \ref{fig:PIAID_scheme}}. Interference alignment
is applied only to the members of the alignment set.
We derive the average SER by taking into account the non-Guassian
residual interference { due to discrete constellation.} Using graph theory, { we transform the combinatorial problem into a linear programming (LP) problem and obtain a low
complexity user set selection algorithm for the PIAID
scheme, which minimizes the asymptotically tight bound for the average
end-to-end SER performance.} Furthermore, using Semi-Definite
Relaxation (SDR) technique
\cite{sdr:2002,sdr:2004,sdr:mimo:2005,sdr:mimo:2005_2}, we propose a
low complexity ID algorithm at the receiver. The SER performance of
the proposed PIAID scheme is shown to have significant gain compared
with various conventional baseline solutions.

{\em Outline}: The rest of this paper is organized as follows. In
Section \ref{sec:model}, we outline the system model and the
proposed PIAID scheme. In Section \ref{sec:pia}, we discuss the
optimization of the IA user selection. In Section
\ref{sec:two_stage}, we derive the average SER by taking into
account the non-Guassian residual interference and propose a low
complexity ID algorithm at the receiver that uses SDR technique. The
numerical simulation results are illustrated in Section
\ref{sec:sim}. Finally we conclude with a brief summary in Section
\ref{sec:con}.

\section{System Model}\label{sec:model}
\subsection{$K$-user Quasi-Static MIMO Interference Channels}
\begin{figure}
\centering
\includegraphics[width = 14cm]{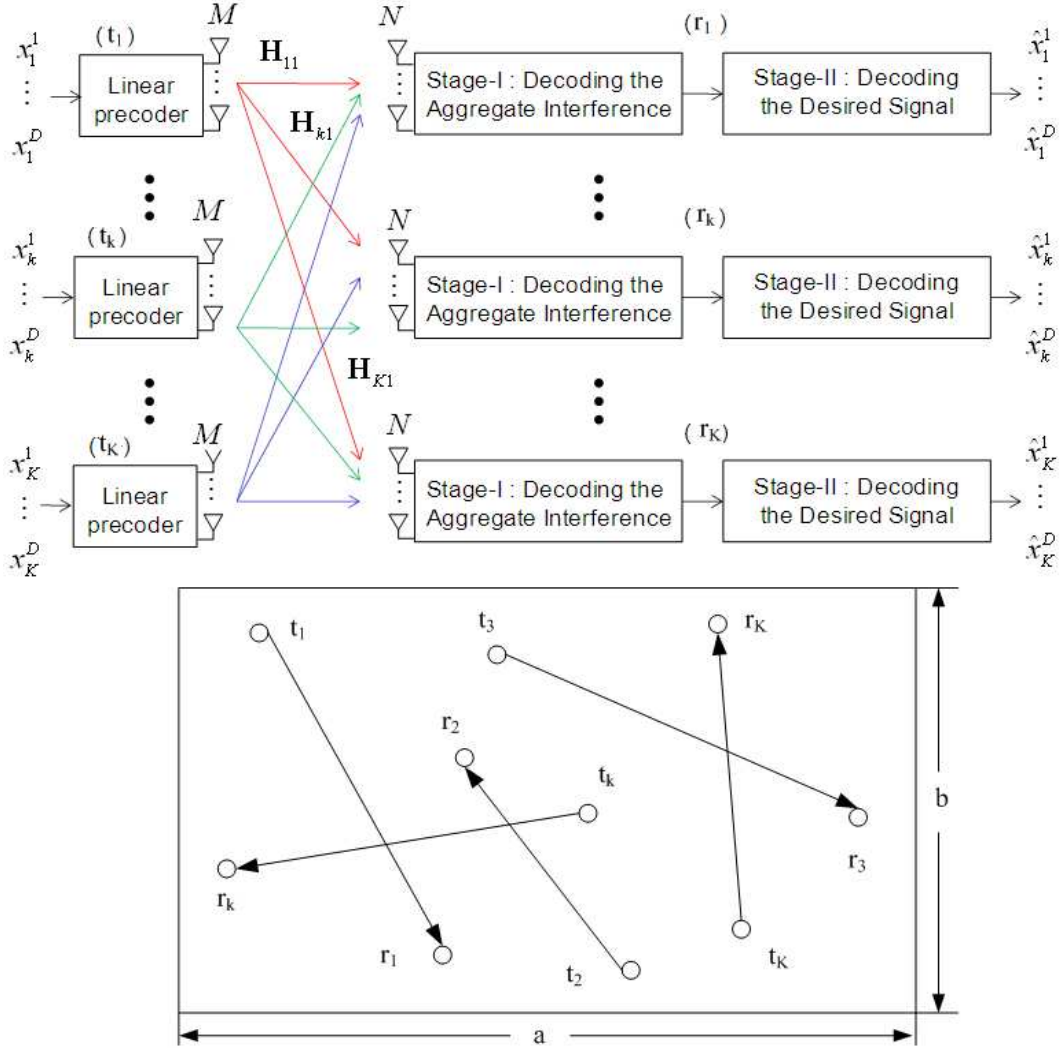}
\caption{Quasi-static $K$-user MIMO complex Gaussian interference
channels. Each $M$-antenna transmitter tries to transmit $D$
independent data streams to its corresponding $N$-antenna receiver.
All the nodes are randomly distributed in the $a\times b$
rectangular area.} \label{fig:system_model}
\end{figure}

We consider $K$-user quasi-static MIMO Gaussian interference
channels as illustrated in Fig. \ref{fig:system_model}.
Specifically, each $M$-antenna transmitter, tries to communicate to
its corresponding $N$-antenna receiver. The channel output at the
$k$-th receive node is described as follows:
\begin{equation}\label{eq:system_model}
\mathbf{y}_k=\sum_{i\in\mathcal{K}}\sqrt{P_iL_{ki}}\mathbf{H}_{ki}\mathbf{\overline{x}}_i+\mathbf{z}_k,
\end{equation}
where $\mathcal{K}=\{1,\cdots,K\}$,
$\mathbf{H}_{ki}\in\mathbb{C}^{N\times M}$ is the MIMO complex
fading coefficients from the $i$-th transmitter to the $k$-th
receiver, $L_{ki}$ is the long term path gain from the $i$-th
transmitter to the $k$-th receiver, and $P_i$ is the average
transmit power of the $i$-th transmitter. In
(\ref{eq:system_model}),
$\mathbf{\overline{x}}_i\in\mathbb{C}^{M\times 1}$ is the complex
signal vector transmitted by transmit node $i$, and
$\mathbf{z}_k\in\mathbb{C}^{N\times 1}$ is the circularly symmetric
Additive White Gaussian Noise (AWGN) vector at receive node $k$. We
assume all noise terms are i.i.d zero mean complex Gaussian with
$\mathbb{E}[\mathbf{z}_k(\mathbf{z}_k)^{H}]=2\mathbf{I}_{N}$.
Furthermore, the assumption on channel model is given as follows:
\begin{assumption}[Assumption on Channel Model { \cite{ITU:1997}}]\label{ass:channel_model}
We assume that the long term path gain is given by $L_{ki}=\omega
d_{ki}^{-\gamma}$, where $d_{ki}$ is the distance between transmit
node $i$ and receive node $k$, $\omega$ is the Log-normal shadow
fading with a standard deviation $\sigma_{\omega}$, and $\gamma$ is
the path loss exponent. Furthermore, we assume that the entries of
$\mathbf{H}_{ki}$ for all $k,i$ are i.i.d. complex Gaussian random
variables given by $[\mathbf{H}_{ki}]_{(n,m)} \sim
\mathcal{CN}(0,1)$ for all $k,i,n,m$, where $[\mathbf{H}_{ki}]_{(n,m)}$ denotes the
$(n,m)^{\text{th}}$ element of $\mathbf{H}_{ki}$. ~\hfill\IEEEQED
\end{assumption}

In this paper, we assume that the $i$-th transmit node transmits
$D\leq\min(M,N)$ independent QPSK data streams $\{x_i^1,\cdots,
x_i^D\}$ to the $i$-th receive node where
$x_i^d\in\mathcal{S}=\{\frac{\sqrt{2}}{2}(1+j),\frac{\sqrt{2}}{2}(1-j),
\frac{\sqrt{2}}{2}(-1+j),\frac{\sqrt{2}}{2}(-1-j)\},\forall
d\in\{1,\cdots,D\}$. Let $\mathbf{v}_i^d$ ($||\mathbf{v}_i^d||=1$, where $||\cdot||$ denotes the Frobenius norm.)
denote the precoder for the $x_i^d$ symbol. Hence, the transmitted
vector at the $i$-th transmitter is given by
$\mathbf{\overline{x}}_i=\sum_{d}\mathbf{v}_i^dx_i^d$.

\subsection{
Partial Interference Alignment and Interference Detection (PIAID)
Scheme}

The proposed PIAID scheme consists of two major components, namely
the {\em Partial Interference Alignment} (PIA) at the transmitters
and the {\em Interference Detection} (ID) at the receivers { as illustrated in Fig. \ref{fig:PIAID_scheme}.}

\subsubsection{Overview of PIA} 
\begin{figure}
\centering
\includegraphics[width = 14cm]{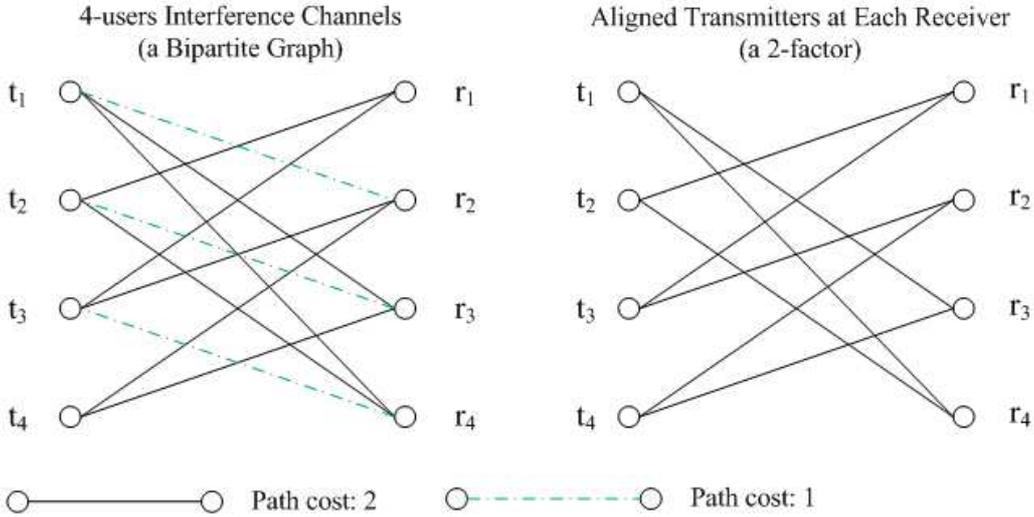}
\caption{ An example of 4-user interference channels where $M=N=2$
and $D=1$. Using the feasibility condition of MIMO IA in
\cite{Feasibility:MIMO:2009}, only two transmitters can be aligned
at each receiver. Specifically, the 4-user interference channels are
represented by a bipartite graph
$\mathcal{B}=(\mathcal{R},\mathcal{T},\mathcal{E})$, where
$\mathcal{R}=\{r_1,\cdots,r_4\}$ is the set of the receive nodes,
$\mathcal{T}=\{t_1,\cdots,t_4\}$ is the set of the transmit nodes,
and $\mathcal{E}=\{[r_k,t_i],\forall k,i\text{ and } k\neq i\}$ is
the set of the edges. Furthermore, the PIA set is given by
$\mathcal{A}=\{\mathcal{A}_1=\{2,3\},\mathcal{A}_2=\{3,4\},\mathcal{A}_3=\{4,1\},\mathcal{A}_4=\{1,2\}
\}$ and the corresponding subset of edges is given by
$\mathcal{E}_s=\{[r_1,t_2],[r_1,t_3],[r_2,t_3],[r_2,t_4],[r_3,t_4],[r_3,t_1],[r_4,t_1],[r_4,t_2]\}$.}
\label{fig:aligned_user}
\end{figure}

PIA is motivated by the
feasibility issue\footnote{ For constant MIMO interference channels, it is not always possible to completely align all the $K-1$ interferers at each receiver.} of the MIMO interference alignment without symbol
extension \cite{Feasibility:MIMO:2009,IA:distributed:2008}. For
instance, it is conjectured in \cite{Feasibility:MIMO:2009} that only when
$K\leq\frac{M+N}{D}-1$, the $K-1$ interfering transmitters can be
aligned at every receiver node. As a result, not all the $K-1$
interfering transmitters can be aligned at every receiver node for
large $K$. Furthermore, existing IA schemes do not consider or
exploit the effects of different path losses between transmit and
receive pairs. When the path loss effects are taken into
consideration, not all transmitters will contribute the same effect
at the receiver and hence, there should be different priority in
determining which nodes should be aligned given the feasibility
constraint. Fig. \ref{fig:aligned_user} illustrates an example of
4-user interference channels where $M=N=2$ and $D=1$. Using the
feasibility condition of MIMO IA in \cite{Feasibility:MIMO:2009},
only two transmitters can be aligned at each receiver. Combining
with the path costs (which depend on the path gains and transmit
power), it is obvious that transmitters 2 and 3 should be aligned at
receiver 1 as indicated in the Fig. \ref{fig:aligned_user}.

{ Motivated by the above example, the proposed PIAID scheme
dynamically selects $\alpha$ transmitters to be aligned at each
receiver node based on the path costs. The index of the aligned
transmitters at each receiver is given by a PIA set with cardinality $\alpha$. Specifically, the PIA set is
defined below.}
\begin{Def}[PIA Set]\label{def:ia_set}
A PIA set is defined as $\mathcal{A}=\{\mathcal{A}_k,\forall k\}$,
where $\mathcal{A}_k=\{k_1,k_2,\cdots,k_{\alpha}\neq k\}$ denotes
the index of aligned transmitters at the receiver $k$ for some
constant $\alpha$. ~\hfill\IEEEQED
\end{Def}

Only the transmit nodes that belong to $\mathcal{A}_k$ will have to
align their transmit signals by choosing the precoders and
equalizers according to the traditional IA requirement\footnote{ Note that
the requirement is equivalent to that used in \cite{Feasibility:MIMO:2009,Tresch:2009}, i.e., choosing the precoders and
equalizers satisfying $\text{rank}((\mathbf{U}_k)^{\dag}\mathbf{H}_{kk}\mathbf{V}_{k} )=D,\forall k$.}
\cite{Feasibility:MIMO:2009,Tresch:2009}
\begin{eqnarray}\label{eq:ia_constraint}
(\mathbf{U}_k)^{\dag}\mathbf{H}_{ki}\mathbf{V}_{i}&=&\mathbf{0},
\forall i\in
\mathcal{A}_k\nonumber\\
(\mathbf{U}_k)^{\dag}\mathbf{H}_{kk}\mathbf{V}_{k} &=&
\text{diag}(\lambda_1,\cdots,\lambda_D),\forall k
\end{eqnarray}
where { $(\cdot)^\dag$ denotes the Hermitian
transpose, $\text{diag}(\lambda_1,\cdots,\lambda_D)$ denotes a diagonal matrix
with diagonal entries $\lambda_1,\cdots,\lambda_D$,}
$\mathbf{U}_k=[\mathbf{u}_k^1,\cdots,\mathbf{u}_k^D]$ are the
$N\times D$ decorrelators at receiver $k$ with
$||\mathbf{u}_k^l||=1,\forall l$, { and
$\mathbf{V}_i=[\mathbf{v}_i^1,\cdots,\mathbf{v}_i^D]$ are the
$M\times D$ precoders at transmitter $i$ with
$||\mathbf{v}_i^l||=1,\forall l$.} { Based on the conjecture in \cite{Feasibility:MIMO:2009}, a sufficient condition\footnote{ From \cite{Feasibility:MIMO:2009}, we know that the total number
of equations for the IA requirement $\{(\mathbf{U}_k)^{\dag}\mathbf{H}_{ki}\mathbf{V}_{i}=\mathbf{0},
\forall i\in
\mathcal{A}_k,\forall k\}$ is $N_e=\sum_{k}|\mathcal{A}_k|D^2=\alpha KD^2$,
and the total number of variables is $N_v=\sum_{k}D(M+N-2D)=KD(M+N-2D)$. When each transmitter is selected by $\alpha$ receivers, the feasibility
condition is simply given by $N_e\leq N_v$ \cite{Feasibility:MIMO:2009}, i.e., $\alpha(\mathcal{A})\leq\min\left( \frac{M+N}{D}-2,K-1\right)$.} for a feasible PIA set $\mathcal{A}$ is given by\cite{Feasibility:MIMO:2009}:}
\begin{equation}\label{eq:ia_feasibility}
\alpha(\mathcal{A})\leq\min\left( \frac{M+N}{D}-2,K-1\right)\text{
and } \sum\nolimits_{k}\mathbf{1}(i\in
\mathcal{A}_k)=\alpha(\mathcal{A}),\forall i
\end{equation}
where $\mathbf{1}(\cdot)$ is the indicator function, $\alpha(\mathcal{A}) = |\mathcal{A}_k|$ is the cardinality of
$\mathcal{A}_k$ (i.e., the number of aligned interferers at each
receiver). The requirement as per (\ref{eq:ia_feasibility}) means
that each transmitter should be selected by $\alpha(\mathcal{A})$
receivers. The IA set $\mathcal{A}$ is a design parameter in the
proposed PIA scheme, and { how to choose the IA set is presented in Section \ref{sec:pia}.}

{
\begin{Rem}[Feasibility Condition]
Note that the feasibility conditions for \eqref{eq:ia_constraint} are still open in the literature. Since the feasibility condition and computation of $\{\mathbf{U}_k,\mathbf{V}_k,\forall k\}$ are not the focus of the paper, we have adopted the results in \cite{Feasibility:MIMO:2009} to derive a sufficient condition \eqref{eq:ia_feasibility}. While this condition restricts the choice of the feasible set $\mathcal{A}$, it introduces graph structure for the optimization w.r.t. $\mathcal{A}$. Furthermore, as shown in Fig. \ref{fig:5users}, the proposed PIA algorithm with condition \eqref{eq:ia_feasibility} has similar performance as the solution obtained by brute-force exhaustive search (without \eqref{eq:ia_feasibility}). ~\hfill\IEEEQED
\end{Rem}

\subsubsection{Overview of ID}
In this paper, we focus on the case where not all the $K-1$ interferers can be aligned at each receivers. As a result, there will be residual interference at every receiver.
The ID processing at the receiver first estimates the aggregate interference signal by using the constellation maps derived from the QPSK inputs. The desired signal is detected after subtracting the estimated aggregate interference.} For
instance, the normalized received signal at the $k$-th receiver is
given by:
\begin{equation}
\label{eq:y_k}
\begin{array}{ll}
&\mathbf{y}_k=\sqrt{P_{k}L_{kk}}\mathbf{H}_{kk}\mathbf{v}_k^lx_k^l
+\underbrace{\sum\nolimits_{d\neq
l}\sqrt{P_{k}L_{kk}}\mathbf{H}_{kk}\mathbf{v}_k^dx_k^d}_{\text{inter-stream
interference}}\\
&\quad\quad\quad\quad+\underbrace{\sum\nolimits_{i\not\in
\mathcal{A}_k;d}\sqrt{P_{i}L_{ki}}\mathbf{H}_{ki}\mathbf{v}_i^dx_i^d}_{\text{non-aligned
interferers}} +\underbrace{\sum\nolimits_{i\in \mathcal{A}_k
;d}\sqrt{P_{i}L_{ki}}\mathbf{H}_{ki}\mathbf{v}_i^dx_i^d}_{\text{aligned
interferers}} + \mathbf{z}_k.
\end{array}
\end{equation}
We adopt linear processing at the receiver and the detection process
for $l$-th data stream at the $k$-th receiver is divided into two
stages, namely the {\em aggregate interference detection stage}
(stage I) and the {\em desired signal detection stage} (stage II).
The two stages are elaborated below:
\begin{itemize}
\item{\bf Stage I Processing:} Using the $l$-th column of
$\mathbf{U}_k$ in (\ref{eq:ia_constraint}), $\mathbf{u}_k^l$, as the
decorrelator, the post-processed signal of the $l$-th stream is
given by:
\begin{eqnarray}
\label{eq:y_k^l_1st}
y_k^l=(\mathbf{u}_k^l)^{\dag}\mathbf{y}_k=\sqrt{P_{k}L_{kk}}H_{kk}^{ll}x_k^l
+\underbrace{\sum_{i\not\in
\mathcal{A}_k;d}\sqrt{P_{i}L_{ki}}H_{ki}^{ld}x_i^d}_{\text{non-aligned
interferers}}+(\mathbf{u}_k^l)^{\dag}\mathbf{z}_k,
\end{eqnarray}
where
$H_{ki}^{ld}=(\mathbf{u}_k^l)^{\dag}\mathbf{H}_{ki}\mathbf{v}_i^d$
is the equivalent channel gain for the $d$-th data stream of
transmitter $i$ at receiver $k$. Note that the inter-stream
interference and the interference contributed by users in the IA set
$\mathcal{A}_k$ are completely eliminated due to the PIA requirement
in (\ref{eq:ia_constraint}). Let { $\mathcal{Q}_k=\{i:P_{i}L_{ki}\geq
P_{k}L_{kk},\forall i\not\in\mathcal{A}_k , i\neq k\}\subseteq\{1,\cdots,K\}$} denotes the
set of {\em strong residual interference}. { Since ID at the receiver is more effective
when the interference is stronger than the desired signal \cite{Symmetric:2008,Real:MIMO:2009}}, the first stage
processing estimates the {\em aggregate strong interference}
$I_k^l=\sum\nolimits_{i\in\mathcal{Q}_k;d}\sqrt{P_{i}L_{ki}}H_{ki}^{ld}x_i^d$
using the following nearest neighbor detection rule.
\begin{Alg}[Stage I Interference Detection Algorithm]\label{alg:stage-I}
Based on the decorrelator output, $y_k^l$, the detected aggregate
strong interference $\hat{I}_k^l$ is given by:
\begin{equation}
(\hat{I}_{k}^l)^* = \argmin_{c\in\mathcal{I}_k^l}|y_k^l-c|,
\end{equation}
{ where $\mathcal{I}_{k}^l
=\big\{\sum\limits_{i\in\mathcal{Q}_k;d}\sqrt{P_{i}L_{ki}}H_{ki}^{ld}s_i^d:
s_i^d\in \mathcal{S}\big\}$ is the set of possible values the
strong interference from $\mathcal{Q}_k$ can take.} ~\hfill\IEEEQED
\end{Alg}

Note that when $\mathcal{Q}_k=\emptyset$, there will be no stage I
decoding for the desired data stream $x_k^l$. In this case, the
proposed PIAID scheme reduces to the conventional receiver with
one-stage decoding.

\item{\bf Stage II Processing:}
The estimated aggregate { strong} interference $(\hat{I}_k^l)^*$
is first subtracted from the decorrelator output $y_k^l$ as
illustrated:
\begin{equation}
\label{eq:y_k^l_2nd} \widetilde{y}_k^l=y_k^l-{
(\hat{I}_k^l)^*}=\sqrt{P_{k}L_{kk}}H_{kk}^{ll}x_k^l+\underbrace{I_k^l-(\hat{I}_k^l)^*+\sum_{i\in
\mathcal{O}_k;d} \sqrt{P_{i}L_{ki}}H_{ki}^{ld}x_i^d}
_{\text{residual interference}}+(\mathbf{u}_k^l)^{\dag}\mathbf{z}_k,
\end{equation}
where { $\mathcal{O}_k=\{i:P_{i}L_{ki}< P_{k}L_{kk},\forall
i\not\in\mathcal{A}_k , i\neq k\}\subseteq\{1,\cdots,K\}$} denotes the set of {\em weak
residual interference}, and obviously we have
$\mathcal{O}_k\bigcup\mathcal{Q}_k\bigcup\mathcal{A}_k\bigcup\{k\}=\mathcal{K}$.
In turn, the desired signal for receiver $k$ is detected based on
$\widetilde{y}_k^l$ using the following algorithm.
\begin{Alg}[Stage II Signal Detection Algorithm]\label{alg:stage-II} The
$l$-th data symbol at the $k$-th receiver $x_k^l$ is detected based
on $\widetilde{y}_k^l$ according to the minimum-distance rule given
by:
\begin{equation}
\Re\{(\hat{x}_{k}^l)^*\} = \left\{\begin{array}{ll}
\frac{\sqrt{2}}{2} & \text{if }
\Re\left\{\frac{\widetilde{y}_k^l}{\sqrt{P_{k}L_{kk}}H_{kk}^{ll}}\right\}\geq 0 \\
-\frac{\sqrt{2}}{2} & \text{else}
\end{array}\right.\quad
\Im\{(\hat{x}_{k}^l)^*\} = \left\{\begin{array}{ll}
\frac{\sqrt{2}}{2} & \text{if }
\Im\left\{\frac{\widetilde{y}_k^l}{\sqrt{P_{k}L_{kk}}H_{kk}^{ll}}\right\}\geq 0 \\
-\frac{\sqrt{2}}{2} & \text{else}
\end{array}\right.
\end{equation}
~\hfill\IEEEQED
\end{Alg}
\end{itemize}

\begin{figure}
\centering
\includegraphics[width = 12cm]{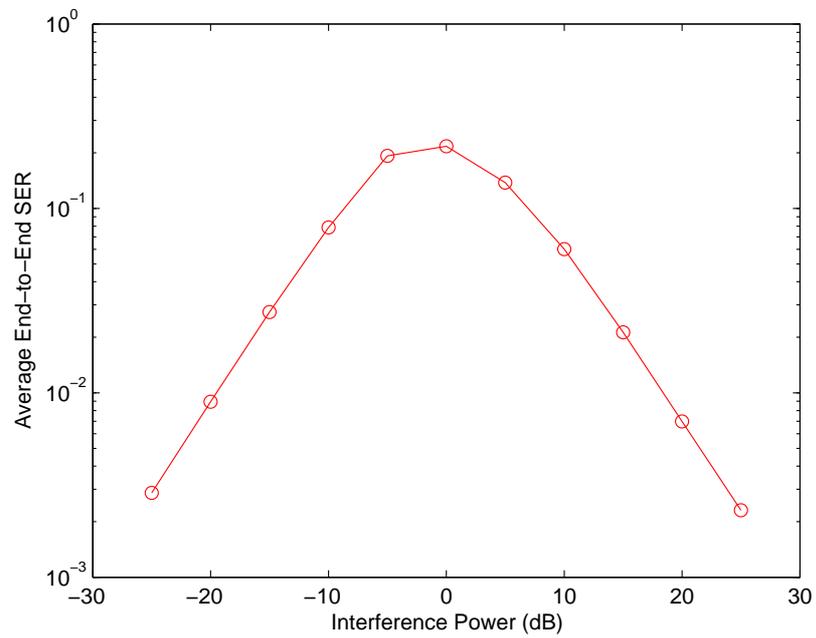}
\caption{Average end-to-end SER performance of the ID detector
versus the interference power at the interference limited regime.
Specifically, the input to the ID detector is given by
$y=H_1x_1+\sqrt{P_2}H_2x_2$, where $x_1\in\mathcal{S}$ is the
desired signal, $x_2\in\mathcal{S}$ is the interference,
$H_1,H_2\sim\mathcal{CN}(0,1)$, and the interference power is
$P_2$. The average end-to-end SER is given by
$\overline{\text{SER}}(P_2)\triangleq\sum_{x_1}\Pr\{\hat{x}_1\neq
x_1|x_1,P_2\}$.}\label{fig:ser_ic}
\end{figure}

Note that the performance of the ID processing depends heavily on
the {\em interference profile}, { which contains the relative power of the residual interference at the receiver}. Fig. \ref{fig:ser_ic} illustrates
the average end-to-end SER performance of the ID detector versus the
{ interference power}. Observe that there is a window of unfavorable
interference power for which the performance of the ID is quite
poor. As such, the user selection of the PIA stage can contribute
significantly to the end-to-end SER performance of the PIAID scheme.
Intuitively, the user set selection $\mathcal{A}$ of PIAID should
not aim at removing the {\em strongest interference}. On the
contrary, it should remove the {\em unfavorable interference}
characterized by the ID stage requirement (similar to Fig.
\ref{fig:ser_ic}). As a result, the PIA and ID processing are
complementary approaches to combat interference and their designs
are tightly coupled together.

\section{Dynamic IA Set Selection in Partial Interference Alignment}\label{sec:pia}
In this section, we shall formulate the dynamic selection of PIA set
as a combinatorial optimization problem, and derive a low complexity
optimal solution using graph theory and linear programming.

\subsection{Problem Formulation}
Due to heterogeneous path losses and transmit powers, interference
links have different contributions to the average end-to-end SER of
the PIAID scheme. We shall first formulate the dynamic PIA set
selection problem using a general cost metric
$\mathcal{C}=\{c_{ki},\forall k,i\}$. In the next section, we shall
obtain a specialized cost metric related to the end-to-end SER of
the PIAID. The PIA set optimization problem is summarized below.
{
\begin{Prob}[MaxPIA Problem]\label{prob:ia_set}
Given a general cost matrix of the interference links
$\mathcal{C}=\{c_{ki},\forall k,i\}$, the MaxPIA prolbem is given
by:
\begin{equation}
\label{eq:max_prob}\text{MaxPIA:}\quad\quad\mathcal{A}^*=\argmax_{\mathcal{A}\in\mathbb{A}}
\sum_{k,i}c_{ki}\mathbf{1}(i\in \mathcal{A}_k),
\end{equation}
where $\mathcal{A}^*$ is the solution to the MaxPIA problem, and $\mathbb{A}$ denotes the collection of all the PIA sets that
satisfies the IA feasibility condition (\ref{eq:ia_feasibility}).
~\hfill\IEEEQED
\end{Prob}
}

\subsection{Optimal Solution of the Dynamic PIA Set Selection}
Optimization problem (\ref{eq:max_prob}) is a constrained
combinatorial optimization problem, which is difficult in general.
Solving problem (\ref{eq:max_prob}) using brute force exhaustive
search has a high complexity of $O(\exp(K))$ and is not viable in
practice. In this section, we shall exploit specific problem
structure and visualize the optimization problem in
(\ref{eq:max_prob}) using graph theory.

We first review some preliminaries on graph theory from
\cite{comb:1998,comb:1988} and the reference therein. A {\em graph}
$\mathcal{G}$ is defined by a pair
$\mathcal{G}=(\mathcal{W},\mathcal{E})$, where $\mathcal{W}$ is a
finite set of {\em nodes} and $\mathcal{E}$ is a finite set of {\em
edges}. Specifically, the nodes in $\mathcal{W}$ are denoted as
$w_1,w_2,\cdots,w_{|\mathcal{W}|}$, and an edge in $\mathcal{E}$
connecting nodes $w_i$ and $w_k$ is denoted as $[w_i,w_k]$. If an
edge $e=[w_1,w_2]\in\mathcal{E}$, then we say that $e$ is {\em incident}
upon $w_1$ (and $w_2$). The {\em degree} of a node $w$ of
$\mathcal{W}$ is the number of edges incident upon $w$. A {\em
bipartite graph} is a graph $\mathcal{B}=(\mathcal{W},\mathcal{E})$
such that $\mathcal{W}$ can be partitioned into two sets,
$\mathcal{W}_{1}$ and $\mathcal{W}_{2}$, and each edge in
$\mathcal{E}$ has one node in $\mathcal{W}_{1}$ and one node in
$\mathcal{W}_{2}$. The bipartite graph is usually denoted by
$\mathcal{B}=(\mathcal{W}_{1},\mathcal{W}_{2},\mathcal{E})$. An
example of a bipartite graph is illustrated in Fig.
\ref{fig:aligned_user}.

In fact, interference networks can be represented by a bipartite
graph $\mathcal{B}=(\mathcal{R},\mathcal{T},\mathcal{E})$, where
$\mathcal{R}=\{r_1,\cdots,r_K\}$ is the set of the receive nodes,
$\mathcal{T}=\{t_1,\cdots,t_K\}$ is the set of the transmit nodes,
and $\mathcal{E}=\{[r_k,t_i],\forall k,i\text{ and } k\neq i\}$ is
the set of the edges. A feasible PIA set $\mathcal{A}$ is equivalent
to a subset of the edges $\mathcal{E}_s$ with the property that the
degree of each receive and transmit node of
$\mathcal{G}_s=(\mathcal{T},\mathcal{R},\mathcal{E}_s)$ is
$\alpha(\mathcal{A})$, and { $\mathcal{G}_s$ is called a {\em
$\alpha$-factor} of graph $\mathcal{G}$.}
\begin{Exam}[Graph Illustration of the PIA set]
Suppose the PIA set is given by
$\mathcal{A}=\{\mathcal{A}_1=\{2,3\},\mathcal{A}_2=\{3,4\},\mathcal{A}_3=\{4,1\},\mathcal{A}_4=\{1,2\}
\}$, the corresponding subset of edges $\mathcal{E}_s$ is given by
$\mathcal{E}_s=\{[r_1,t_2],[r_1,t_3],[r_2,t_3],[r_2,t_4],[r_3,t_4],[r_3,t_1],[r_4,t_1],[r_4,t_2]\}$
as illustrated in Fig. \ref{fig:aligned_user}. ~\hfill\IEEEQED
\end{Exam}

Let $c_{ki}$ denote the cost of edge $[r_k,t_i]\in\mathcal{E}$.
Problem \ref{prob:ia_set} is equivalent to finding a $\alpha$-factor of
$\mathcal{G}$ with the largest sum of costs. { Hence, the MaxPIA problem in \eqref{eq:max_prob} is similar to a matching problem (finding a ``best'' 1-factor of graph $\mathcal{G}$) on a bipartite graph},
and exploiting this equivalence we shall derive a low complexity
optimal solution. Let $\mathbf{e}=\{e_{ki},\forall k,i\}$ be a set
of variables. If the edge $[r_k,t_i]$ is included in the $\alpha$-factor
(i.e. the transmit node $i$ is chosen as one of the aligned
interferers at receive node $k$) then $e_{ki}=1$, otherwise
$e_{ki}=0$. As a result, the problem \ref{prob:ia_set} is equivalent
to
\begin{equation}
\label{eq:sub_opt} \{e_{ki}^*\}=\left\{
\begin{array}{cl}
\argmax_{e_{ki}} & \sum_{k,i}c_{ki}e_{ki} \\
\text{s.t} & \sum_{i}e_{ki} = \alpha, \forall k\\
&\sum_{k}e_{ki} = \alpha, \forall i\\
&e_{ki} \in \{0,1\}, \forall k,i
\end{array}\right.
\end{equation}
where $\alpha=\min\left(\frac{M+N}{D}-2,K-1\right)$.

The above problem is a non-convex problem due to the non-convex
constraint $e_{ki} \in \{0,1\}$. To get a low complexity solution,
we first relax the constraints $e_{ki} \in \{0,1\}$ to $0\leq e_{ki}
\leq 1$. As a result, (\ref{eq:sub_opt}) becomes a standard LP problem, which can be solved efficiently by the
well known simplex algorithm\cite{comb:1988,comb:1998}. The
following Lemma summarizes the optimality of this relaxation.
\begin{Lem}[Optimality of the LP Relaxation]\label{lem:lp}
The optimal solution of the LP relaxation problem is also the
optimal solution of (\ref{eq:sub_opt}), i.e. $\widetilde{e}_{ki}^*
\in \{0,1\}$ where $\widetilde{e}_{ki}^*$ is the optimal solution of
the LP relaxed problem. ~\hfill\IEEEQED
\end{Lem}
\begin{proof}
Please refer to Appendix \ref{app:lem:lp}.
\end{proof}

\section{SER Analysis and Low Complexity ID Processing}\label{sec:two_stage}
In this section, we first derive the average SER of the
PIAID scheme for a given PIA set $\mathcal{A}$ and the path gains
$\{L_{ki}P_{i},\forall k,i\}$. Based on the SER results, we obtain
an equivalent cost metric for the PIA set selection optimization
problem in (\ref{eq:sub_opt}), which is order-optimal w.r.t. the
average end-to-end SER. Finally, we propose a low complexity ID
algorithm using SDR technique.

\subsection{SER Analysis of ID with Non-Gaussian Residual Interference} Unlike standard SER analysis in existing
literature \cite{Simon:2005,proakis:2001}, a key challenge of SER
analysis in the PIAID scheme is that the residual interference in
the stage I and stage II ID processing are non-Gaussian due to the
discrete constellation inputs. As a result, we focus on deriving an
{ asymptotically tight} SER expression for the interference dominated regime.
Theorem \ref{thm:ser} summarizes our main results.
\begin{Thm}[Average SER of the Two Stage PIAID
Processing]\label{thm:ser} For a given PIA set $\mathcal{A}$, the
SER of the $l$-th data stream at the $k$-th receiver in the
interference limited regime of the PIAID scheme is given by
\begin{equation}\label{eq:ser}
\begin{array}{lll}
\overline{\text{SER}}_k^l(\mathcal{A})&\triangleq&
\mathbb{E}_{\mathcal{H}}\left[\text{SER}_k^l(\mathcal{A},\mathcal{H})\right]\triangleq
\mathbb{E}_{\mathcal{H}}\left[
\sum_{x_k^l}\Pr\{x_k^l\}\Pr\{\hat{x}_k^l\neq
x_k^l|x_k^l,\mathcal{A},\mathcal{H}\}\right]\\
&=& { \ThetaL\left(\sum_{i\in\mathcal{Q}_k}\frac{P_{k}L_{kk}}{P_{i}L_{ki}}+
\sum_{i\in\mathcal{O}_k}\frac{P_{i}L_{ki}}{P_{k}L_{kk}}\right)},
\end{array}
\end{equation}
where $\mathcal{H}=\{\mathbf{H}_{ki},\forall k,i\}$.
{ $g(x)=\ThetaL(f(x))$ denotes $\limsup_{x\to0}\frac{g(x)}{f(x)}\leq C$ and $\liminf_{x\to0}\frac{g(x)}{f(x)}\geq c$ for some constants $C\geq c\geq 0$.} ~\hfill\IEEEQED
\end{Thm}
\begin{proof}
Please refer to the Appendix \ref{app:thm:ser}.
\end{proof}
{
\begin{Rem}In this paper, the precoders and decorrelators $\{\mathbf{V}_k,\mathbf{U}_k,\forall k\}$ are determined based on the IA requirements in \eqref{eq:ia_constraint}. Hence, they are only dependent on the channels in the set $\mathcal{A}$ for which interference is aligned. Since the interference from these channels involved are nulled, the remaining interference has a random channel matrix even though it is now projected on the space $\mathbf{U}_k$. Furthermore, the SER in \eqref{eq:ser} is averaged over realizations of the channels and noise.  ~\hfill\IEEEQED
\end{Rem}
}
\begin{Rem}[Interpretation of Theorem \ref{thm:ser}] The result in (\ref{eq:ser})
indicates that the SER of the PIAID scheme favors either very strong
or very weak residual interference. In other words, there is always
an unfavorable {\em window of residual interferences} as illustrated
in Fig. \ref{fig:ser_ic}. The role of PIA is to eliminate these
unfavorable windows of interferences so that the ID processing is
given a more favorable interference profile.  ~\hfill\IEEEQED
\end{Rem}

{ Motivated by Theorem \ref{thm:ser}}, we set the interference
cost metric in the PIA set optimization problem as
\begin{equation}\label{eq:cost_matric}
c_{ki}=\left\{\begin{array}{ll} -C & \text{if } i=k\\
 \frac{P_{k}L_{kk}}{P_{i}L_{ki}} & \text{else if }
P_{i}L_{ki}\geq P_{k}L_{kk}\\
\frac{P_{i}L_{ki}}{P_{k}L_{kk}} & \text{otherwise}
\end{array}\right.,\forall k\in\mathcal{K}
\end{equation}
where $C>0$ is a large constant {(a sufficiently large $C$ can be chosen as:
$C>\sum_{k,i\neq k}|c_{ki}|$)}. Based on these interference cost
metrics, the PIA set selection solutions solved by the LP relaxation
is order-optimal\footnote{ It can be observed from the following fact. From Theorem \ref{thm:ser}, we have
$\mathcal{A}^*=\argmin_{\mathcal{A}\in\mathbb{A}}\ThetaL
\left(\sum_{k,i\in\mathcal{Q}_k}\frac{P_{k}L_{kk}}{P_{i}L_{ki}}+
\sum_{k,i\in\mathcal{O}_k}\frac{P_{i}L_{ki}}{P_{k}L_{kk}}\right)
=\argmin_{\mathcal{A}\in\mathbb{A}}\ThetaL
\Big(\sum_{k,i\neq
k}\frac{P_{k}L_{kk}}{P_{i}L_{ki}}\mathbf{1}(P_{i}L_{ki}\geq
P_{k}L_{kk}) -
\sum_{k,i\in\mathcal{A}_k}\frac{P_{k}L_{kk}}{P_{i}L_{ki}}
\mathbf{1}(P_{i}L_{ki}\geq P_{k}L_{kk})+\sum_{k,i\neq
k}\frac{P_{i}L_{ki}}{P_{k}L_{kk}}\mathbf{1}(P_{i}L_{ki}<
P_{k}L_{kk}) -
\sum_{k,i\in\mathcal{A}_k}\frac{P_{i}L_{ki}}{P_{k}L_{kk}}
\mathbf{1}(P_{i}L_{ki}< P_{k}L_{kk}) \Big)
=\argmax_{\mathcal{A}\in\mathbb{A}}\ThetaL\Big(
\sum_{k,i\in\mathcal{A}_k}\frac{P_{k}L_{kk}}{P_{i}L_{ki}}
\mathbf{1}(P_{i}L_{ki}\geq P_{k}L_{kk}) +
\sum_{k,i\in\mathcal{A}_k}\frac{P_{i}L_{ki}}{P_{k}L_{kk}}
\mathbf{1}(P_{i}L_{ki}< P_{k}L_{kk}) \Big)$. Hence, the $\Big(\sum_{k,i\in\mathcal{Q}_k}\frac{P_{k}L_{kk}}{P_{i}L_{ki}}+
\sum_{k,i\in\mathcal{O}_k}\frac{P_{i}L_{ki}}{P_{k}L_{kk}}\Big)$ is an asymptotically tight bound for SER when $P_kL_{kk} \gg P_iL_{ki}$ or $P_kL_{kk} \ll P_iL_{ki}$ for all $\{k,i\}$, and an order-optimal solution means that it minimizes the asymptotically tight bound for the SER.} w.r.t. the following problem:
\begin{equation}
\mathcal{A}^*=\argmin_{\mathcal{A}\in\mathbb{A}}\sum_{k,l}
\overline{\text{SER}}_k^l(\mathcal{A}).
\end{equation}

\subsection{Low Complexity ID}
Note that the complexity of decoding algorithm \ref{alg:stage-I} in
the stage I processing is exponential w.r.t. the cardinality of the
set of strong residual interferences, i.e., $|\mathcal{Q}_k|$. Using
the SDR technique
\cite{sdr:2002,sdr:2004,sdr:mimo:2005,sdr:mimo:2005_2}, we shall
derive a low complexity ID algorithm, which has polynomial
complexity w.r.t. $|\mathcal{Q}_k|$. The SDR technique has been
widely used in multiuser detection\cite{sdr:2002,sdr:2004} and MIMO
systems \cite{sdr:mimo:2005,sdr:mimo:2005_2} to derive low
complexity suboptimal detectors. It has been shown that the SDR
detector can provide better performance compared with other
suboptimal detectors.

To utilize the SDR technique, we first simplify equation
(\ref{eq:y_k^l_1st}) as follows:
\begin{equation}\label{eq:sdr_y_k_l}
y_k^l=
(\mathbf{h}_{\mathcal{Q}})^{T}\mathbf{x}_{\mathcal{Q}}+\underbrace{
\sqrt{P_{k}L_{kk}}H_{kk}^{ll}x_k^l +\sum_{i\in
\mathcal{O}_k;d}\sqrt{P_{i}L_{ki}}H_{ki}^{ld}x_i^d+
(\mathbf{u}_k^l)^{\dag}\mathbf{z}_k}_{n_0},
\end{equation}
where $\mathbf{x}_{\mathcal{Q}}=[x_{i_1}^1,\cdots,x_{i_1}^D,
x_{i_2}^1,\cdots,x_{i_Q}^D]^{T}\in\mathcal{S}^{ DQ\times 1}$ is { $DQ$ interference symbols}, and
$Q=|\mathcal{Q}_k|$ is the cardinality of $\mathcal{Q}_k$.
$\mathbf{h}_{\mathcal{Q}}=[\sqrt{P_{i_1}L_{ki_1}}H_{ki_1}^{l1},\cdots,\sqrt{P_{i_1}L_{ki_1}}H_{ki_1}^{lD},
\sqrt{P_{i_2}L_{ki_2}}H_{ki_2}^{l1},\cdots,\sqrt{P_{i_Q}L_{ki_Q}}H_{ki_Q}^{lD}]^{T}\\\in\mathbb{C}^{
DQ\times 1}$ { is the channel gain for the interference symbols $\mathbf{x}_{\mathcal{Q}}$}.
Furthermore, the real valued form of (\ref{eq:sdr_y_k_l}) can be
expressed as:
\begin{equation}\label{eq:sdr_real_channel}
\underbrace{\left[\Re\{y_k^l\} \atop
\Im\{y_k^l\}\right]}_{\mathbf{y}_{R}}=\underbrace{\left[\Re\{(\mathbf{h}_{\mathcal{Q}})^{T}\}\quad
-\Im\{(\mathbf{h}_{\mathcal{Q}})^{T}\} \atop
\Im\{(\mathbf{h}_{\mathcal{Q}})^{T}\}\quad\quad
\Re\{(\mathbf{h}_{\mathcal{Q}})^{T}\}
\right]}_{\mathbf{H}_{R}}\underbrace{\left[\Re\{\mathbf{x}_{\mathcal{Q}}\}
\atop
\Im\{\mathbf{x}_{\mathcal{Q}}\}\right]}_{\mathbf{x}_{R}}+\underbrace{\left[\Re\{n_0\}
\atop \Im\{n_0\}\right]}_{\mathbf{n}_{R}},
\end{equation}
where $\mathbf{y}_{R}\in\mathbb{R}^{2\times1}$,
$\mathbf{H}_{R}\in\mathbb{R}^{2\times2DQ}$,
$\mathbf{x}_{R}\in\{\pm1\}^{2DQ\times1}$ and
$\mathbf{n}_{R}\in\mathbb{R}^{2\times1}$. Decoding algorithm
\ref{alg:stage-I} in stage I processing is equivalent to
\begin{equation}
\label{eq:sdr_prob} (\mathbf{x}_{R})^* =
\argmin_{\mathbf{x}_{R}\in\{\pm1\}^{2DQ\times1}}||\mathbf{y}_{R}-\mathbf{H}_{R}\mathbf{x}_{R}||,
\end{equation}
such that { the detected aggregate strong interference}
$(\hat{I}_{k}^l)^*=(\mathbf{h}_{\mathcal{Q}})^{T}(\mathbf{x}_{\mathcal{Q}})^*$,
where $(\mathbf{x}_{\mathcal{Q}})^*$ is determined from
$(\mathbf{x}_{R})^*$ as indicated in (\ref{eq:sdr_real_channel}).
The above problem (\ref{eq:sdr_prob}) can be equivalently expressed
as
\begin{equation}\label{eq:sdr_org}
\mathbf{s}^* = \left\{
\begin{array}{cl}
\argmin\limits_{\mathbf{s}}&\text{trace}(\mathbf{WS})\\
\text{s.t} & \text{diag}(\mathbf{S})=\mathbf{I}_{2DQ+1}\\
&\mathbf{S}=\mathbf{s}\mathbf{s}^T
\end{array}
\right.
\end{equation}
where $(\cdot)^T$ denotes the transpose,
\begin{equation}\label{eq:sdr_W}
\mathbf{W}=\left[\begin{array}{cc} (\mathbf{H}_{R})^T\mathbf{H}_{R}
& -(\mathbf{H}_{R})^T\mathbf{y}_{R} \\
-(\mathbf{y}_{R})^T\mathbf{H}_{R} & 0 \end{array} \right]\quad
\text{ and }\quad \mathbf{s}=\left[\mathbf{x}_{R} \atop 1\right].
\end{equation}

By means of SDR, we relax the constraint
$\mathbf{S}=\mathbf{s}\mathbf{s}^T$ by $\mathbf{S}\succeq0$ { (i.e., $\mathbf{S}\in\mathbb{C}^{n\times n}$ is
positive-semidefinite)}, and
problem (\ref{eq:sdr_org}) degenerates into the following
Semi-Definite Program (SDP) that can be solved efficiently in
$O(Q^{3.5})$
time\cite{sdr:2002,sdr:2004,sdr:mimo:2005,sdr:mimo:2005_2}, e.g.,
using the interior-point optimization technique \cite{Convex:2004}:
\begin{equation}\label{eq:sdr_sdp}
\mathbf{S}^* = \left\{
\begin{array}{cl}
\argmin\limits_{\mathbf{s}}&\text{trace}(\mathbf{WS})\\
\text{s.t} & \text{diag}(\mathbf{S})=\mathbf{I}_{2DQ+1}\\
&\mathbf{S}\succeq 0.
\end{array}
\right.
\end{equation}

If the optimal value $\mathbf{S}^*$ of the above problem
(\ref{eq:sdr_sdp}) is rank one, then the relaxation is tight, and
the optimal solution of the $(\mathbf{x}_R)^*$ is given
by\cite{sdr:2002,sdr:2004,sdr:mimo:2005}:
\begin{equation}\label{eq:sdr_xr}
[(\mathbf{x}_R)^*]_{(n,1)} = f_{\mathbf{x}}\left(\mathbf{r},n\right)
=\left\{\begin{array}{ll} 1 & \text{ if }
\frac{[\mathbf{r}]_{(n,1)}}{[\mathbf{r}]_{(|\mathbf{r}|,1)}} \geq 0 \\
-1 & \text{ else}
\end{array}\right., \quad \forall n\in\{1,\cdots,|\mathbf{r}|-1\}
\end{equation}
where $\mathbf{r}$ is the eigenvector of $\mathbf{S}^*$ associated
with the only one non-zero eigenvalue.

On the other hand, if $\mathbf{S}^*$ is not rank one, then we shall
approximate $(\mathbf{x}_R)^*$ based on $\mathbf{S}^*$.
Specifically, there are a few standard techniques to determine
$(\mathbf{x}_R)^*$, e.g., {\em Randomization}, {\em Rank-1
approximation} and {\em Dominant eigenvector
approximation}\cite{sdr:2002,sdr:2004,sdr:mimo:2005}.

\begin{figure}
\centering
\includegraphics[width = 10cm]{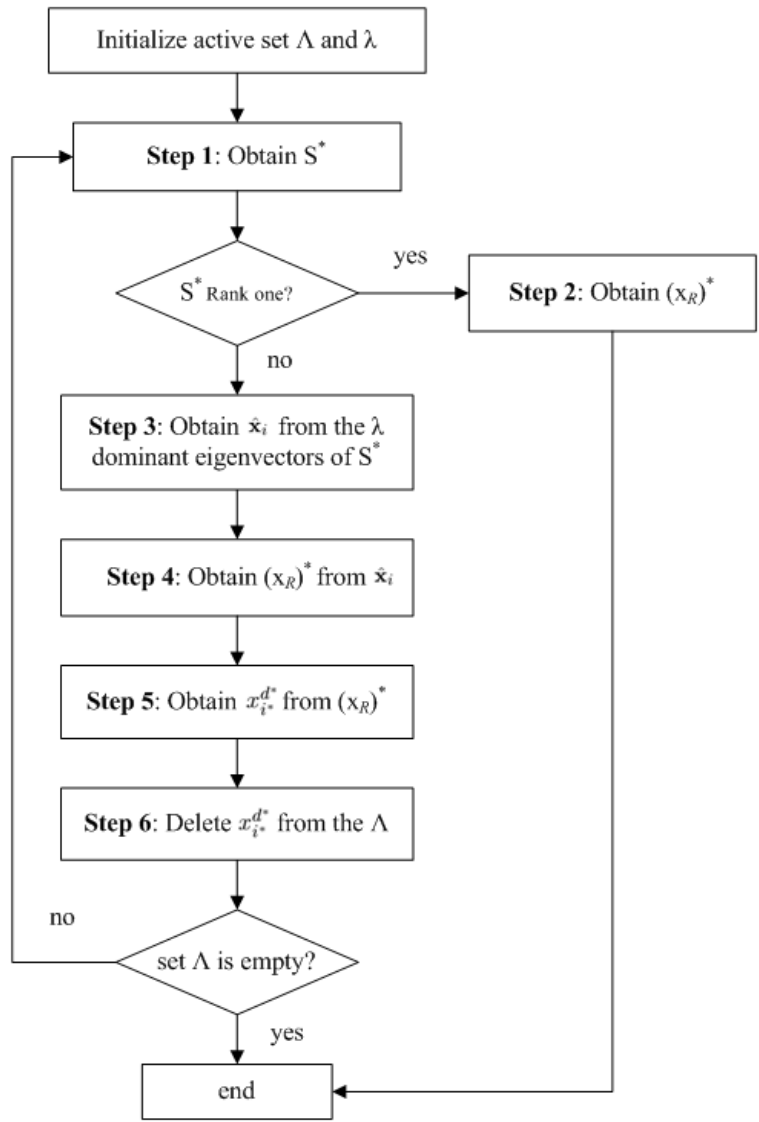}
\caption{Illustration of Algorithm \ref{alg:sdr-sic} (SDR-SID
Algorithm).}\label{fig:sdr_sic}
\end{figure}

To further improve the quality of the approximation, we propose a
SDR-SID algorithm based on the dominant eigenvector approximation as
follows (and illustrated in Fig. \ref{fig:sdr_sic}).
\begin{Alg}[SDR-SID Algorithm]\label{alg:sdr-sic}
\end{Alg}
\begin{itemize}
\item{\bf Step 0:} Set active set $\Lambda=\{x_{i_1}^1,\cdots,x_{i_1}^D,
x_{i_2}^1,\cdots,x_{i_Q}^D\}$ that contains {\em all} the decoding
data streams, and the cardinality is $\lambda=|\Lambda|=DQ$.
\item[] {\bf Repeat}
\item{\bf Step 1:} According to the active set $\Lambda$, solve optimization problem (\ref{eq:sdr_sdp})
to obtain $\mathbf{S}^*$.
\item{\bf Step 2:} If $\mathbf{S}^*$ is rank one, determine
$(\mathbf{x}_R)^*$ from (\ref{eq:sdr_xr}) and terminate.
\item{\bf Step 3:} Extract the $\lambda$ dominant
eigenvectors of $\mathbf{S}^*$,
$\{\mathbf{r}_1,\cdots,\mathbf{r}_{\lambda}\}$. Compute
$[\hat{\mathbf{x}}_i]_{(n,1)} =
f_{\mathbf{x}}\left(\mathbf{r}_i,n\right)$ from (\ref{eq:sdr_xr})
$\forall i\in\{1,\cdots,\lambda\}$ and $\forall
n\in\{1,\cdots,2\lambda\}$.
\item{\bf Step 4:} Compute $\mathbf{S}_i=\left[\hat{\mathbf{x}}_{i}, 1\right]^T\left[\hat{\mathbf{x}}_{i},
1\right]$, $\forall i\in\{1,\cdots,\lambda\}$. Choose
$(\mathbf{x}_R)^*=\hat{\mathbf{x}}_{i^*}$, where
$i^*=\argmin_{i\in\{1,\cdots,\lambda\}}
\text{trace}(W\mathbf{S}_i)$.
\item{\bf Step 5:} Determine $x_{i^*}^{d^*}$ from
$(\mathbf{x}_R)^*$,  where $\{i^*,d^*\}=\argmax_{i,d}
|\sqrt{P_{i}L_{ki}}H_{ki}^{ld}|\mathbf{1}(x_i^d\in\Lambda)$.
\item{\bf Step 6:} Set
$y_k^l=y_k^l-\sqrt{L_{ki^*}}H_{ki^*}^{ld^*}x_{i^*}^{d^*}$, delete
$x_{i^*}^{d^*}$ from the active set $\Lambda$ and set
$\lambda=\lambda-1$.
\item[] {\bf Until} the active set $\Lambda$ is empty. ~\hfill\IEEEQED
\end{itemize}
\begin{Exam}[Illustration of Algorithm \ref{alg:sdr-sic}] Suppose
$y_k^l$ in (\ref{eq:sdr_y_k_l}) is given by
$y_k^l=\sqrt{P_{1}L_{k1}}H_{k1}^{l1}x_1^1+\sqrt{P_{2}L_{k2}}H_{k2}^{l2}x_2^1+n_0$
with
$|\sqrt{P_{2}L_{k2}}H_{k2}^{l2}|>|\sqrt{P_{1}L_{k1}}H_{k1}^{l1}|$.
The details of the implementation of Algorithm \ref{alg:sdr-sic} are
given below.
\begin{itemize}
\item{\bf Step 0:} Set active set $\Lambda=\{x_1^1,x_2^1\}$ and $\lambda=2$.
\item{\bf Step 1:} Suppose $\mathbf{S}^*$ is not rank
one. Go to step 3.
\item{\bf Step 3:} Extract the 2 dominant
eigenvectors of $\mathbf{S}^*$, $\{\mathbf{r}_1,\mathbf{r}_{2}\}$,
and obtain $\{\hat{\mathbf{x}}_1,\hat{\mathbf{x}}_2\}$ from
(\ref{eq:sdr_xr}).
\item{\bf Step 4:} Suppose
$\text{trace}(W\mathbf{S}_1)<\text{trace}(W\mathbf{S}_2)$, we choose
$(\mathbf{x}_R)^*=\hat{\mathbf{x}}_{1}$.
\item{\bf Step 5:} Since
$|\sqrt{P_{2}L_{k2}}H_{k2}^{l2}|>|\sqrt{P_{1}L_{k1}}H_{k1}^{l1}|$,
we determine $x_{2}^{1}$ from $(\mathbf{x}_R)^*$, i.e.,
$x_{2}^{1}=[(\mathbf{x}_R)^*]_{(2,1)}+j[(\mathbf{x}_R)^*]_{(4,1)}$,
which by definition $\mathbf{x}_{\mathcal{Q}}=[x_1^1,x_2^1]$.
\item{\bf Step 6:} Set
$y_k^l=y_k^l-\sqrt{P_{2}L_{k2}}H_{k2}^{l1}x_{2}^{1}=\sqrt{P_{1}L_{k1}}H_{k1}^{l1}x_1^1+n_0$,
$\Lambda=\{x_1^1\}$ and $\lambda=1$. Repeat from step 1 to obtain
$x_1^1$. ~\hfill\IEEEQED
\end{itemize}
\end{Exam}

Algorithm \ref{alg:sdr-sic} is motivated from the intuition that the
error probability of decoding symbol $x_i^d, \forall
i\in\mathcal{Q}_k,\forall d$ is small if its channel gain
$\sqrt{P_{i}L_{ki}}H_{ki}^{ld}$ is large. Note that the complexity
of Algorithm \ref{alg:sdr-sic} is mainly determined by the
complexity of solving the optimization problem (\ref{eq:sdr_sdp}) in
the step 1. Since the complexity for step 1 is only in $O(Q^{3.5})$
time\cite{sdr:2002,sdr:2004,sdr:mimo:2005,sdr:mimo:2005_2}, the
overall complexity of Algorithm \ref{alg:sdr-sic} is $O(Q^{4.5})$.
Furthermore, it can be easily generalized to other approximation
techniques by simply modifying the way to determine
$\hat{\mathbf{x}}_i$ in step 3. Finally, using
simulation, we illustrate that the average end-to-end SER performance of the low
complexity SDR-SID algorithm is similar to the performance of
Algorithm \ref{alg:stage-I}.


\section{Simulation Results and Discussions}\label{sec:sim}
In this section, we evaluate the performance of the proposed PIAID
scheme via numerical simulations. In particular, we compare the
performance of the proposed schemes against various baseline
schemes:
\begin{itemize}
\item{\bf Proposed Scheme 1: PIAID with Algorithm \ref{alg:stage-I} (PIAID Alg1)}
\begin{itemize}
\item The PIA set optimization stage tries to align the unfavorable interference links
by setting the interference cost metric according to
(\ref{eq:cost_matric}).
\item ID processing is adopted at each of the
$K$ receivers after PIA. Specifically, Algorithm \ref{alg:stage-I}
and \ref{alg:stage-II} are applied at stage I and stage II
processing, respectively.
\end{itemize}

\item{\bf Proposed Scheme 2: PIAID with SDR-SID (PIAID Alg3)}
\begin{itemize}
\item The PIA set optimization stage tries to align the unfavorable interference links
by setting the interference cost metric according to
(\ref{eq:cost_matric}).
\item ID processing is adopted at each of the
$K$ receivers after PIA. Specifically, Algorithm \ref{alg:sdr-sic}
and \ref{alg:stage-II} are applied at stage I and stage II
processing, respectively.
\end{itemize}

%

\item{\bf Baseline 1: Randomized PIA (Randomized PIA)}
\begin{itemize}
\item The PIA set $\mathcal{A}$ is chosen randomly from $\mathbb{A}$, i.e., the collection of all the PIA sets that
satisfies the IA feasibility condition (\ref{eq:ia_feasibility}).
\item ID processing is adopted at each of the
$K$ receivers after PIA. Specifically, Algorithm \ref{alg:stage-I}
and \ref{alg:stage-II} are applied at stage I and stage II
processing, respectively.
\end{itemize}

\item{\bf Baseline 2: Iterative interference alignment (Iterative IA)}\cite[Algorithm
1]{IA:distributed:2008}
\begin{itemize}
\item Alternating optimization is utilized to minimize the weighted sum leakage interference.
\item Conventional one-stage decoding is adopted at each of the
$K$ receivers by treating all the interference as noise.
\end{itemize}

\item{\bf Baseline 3: Maximizing SINR (Max SINR)} \cite[Algorithm
2]{IA:distributed:2008}
\begin{itemize}
\item Alternating optimization is utilized to maximize the SINR at the receivers.
\item Conventional one-stage decoding is adopted at each of the
$K$ receivers by treating all the interference as noise.
\end{itemize}
{
\item{\bf Baseline 4: Maximizing Sum-Rate (Max Sum-Rate)} \cite{maximum:sum:rate}
\begin{itemize}
\item A gradient ascent approach combined with the alternating optimization is utilized to maximize the sum-rate of the receivers.
\item Conventional one-stage decoding is adopted at each of the
$K$ receivers by treating all the interference as noise.
\end{itemize}

\item{\bf Baseline 5: Minimizing Mean Square Error (Min MSE)} \cite{alter:VT:2011}
\begin{itemize}
\item A joint design to minimize the sum of the MSE of the receivers.
\item Conventional one-stage decoding is adopted at each of the
$K$ receivers by treating all the interference as noise. ~\hfill\IEEEQED
\end{itemize}
}
\end{itemize}

{ In the simulations, all the transmit and receive nodes are assumed randomly distributed in a
$2\text{km}\times 1\text{km}$ rectangular area as shown in Fig.
\ref{fig:system_model}. The channel model is given by Assumption
\ref{ass:channel_model}. Specifically, we set the log-normal
shadowing standard deviation as $\sigma_{\omega}=12$ dB and the path
loss exponent as $\gamma=6$ as per ITU-R recommendation
M.1225\cite{ITU:1997}. Each transmitter delivers a single stream
$(D=1)$ of QPSK symbols. The transmit power of each node is assumed
to be the same.}



\subsection{Performance w.r.t. Receive $E_s/N_0$}
\begin{figure}
\centering
\includegraphics[width = 14cm]{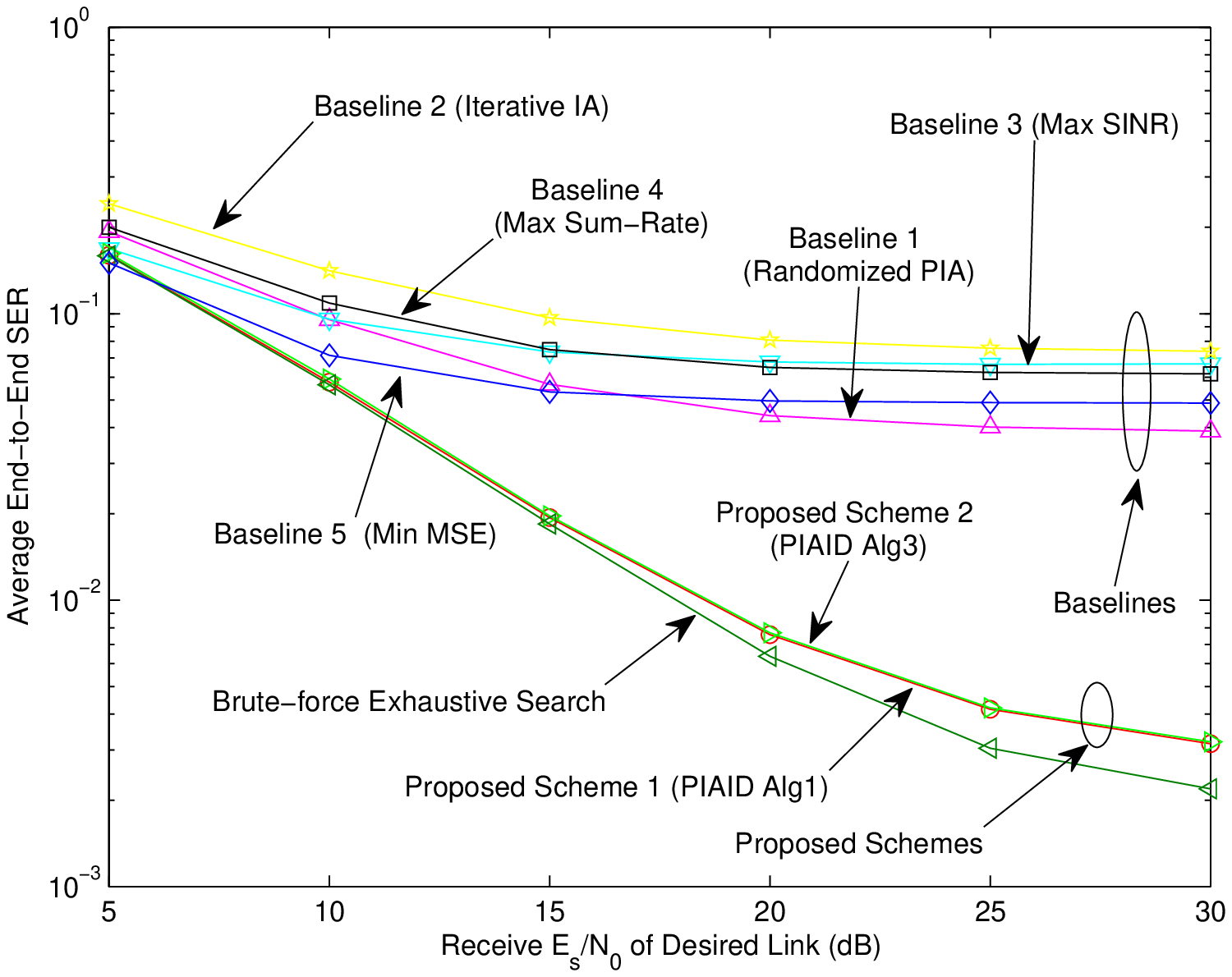}
\caption{ Average end-to-end SER performance versus the receive
$E_s/N_0$ (dB). The setup is given by $K = 5$ (number of users),
$\{M=3,N=2\}$ (number of transmit and receive antennas), $D = 1$
(number of data stream), and $\alpha=3$ (number of aligned users for
feasible interference alignment). }\label{fig:5users}
\end{figure}

\begin{figure}
\centering
\includegraphics[width = 14cm]{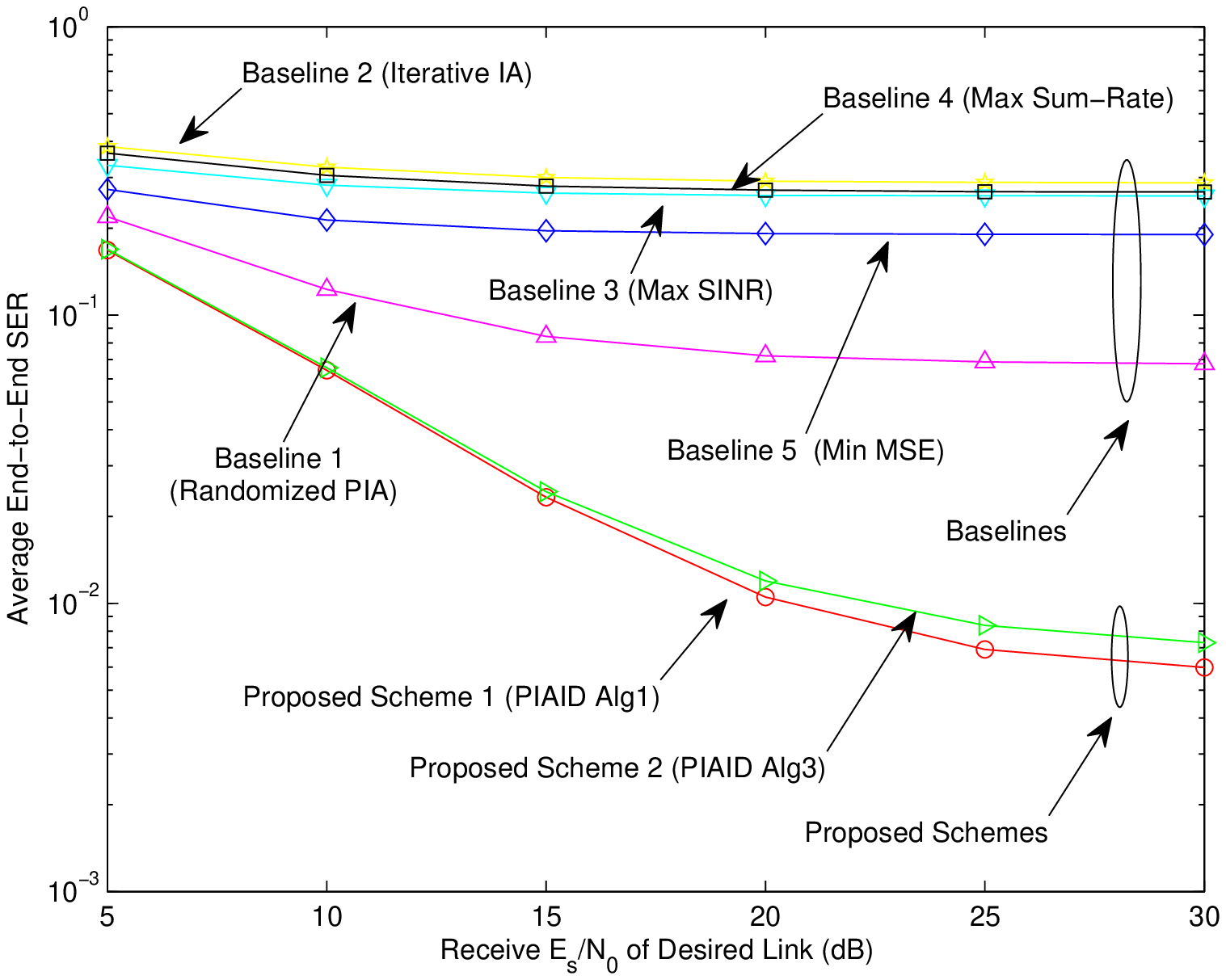}
\caption{ Average end-to-end SER performance versus the receive
$E_s/N_0$ (dB). The setup is given by $K = 6$ (number of users),
$\{M=3,N=2\}$ (number of transmit and receive antennas), $D = 1$
(number of data stream), and $\alpha=3$ (number of aligned users for
feasible interference alignment). }\label{fig:6users}
\end{figure}

Fig. \ref{fig:5users} and Fig. \ref{fig:6users} illustrate the
average end-to-end SER performance per data stream versus receive
$E_s/N_0$ (dB) of the desired link for $K=5$ and $K=6$ respectively.
The number of transmit and receive antennas is given by
$\{M=3,N=2\}$. The number of aligned users for the feasible
interference alignment is $\alpha=3$. { The average SER performance is evaluated with $10^7$ realizations of noise, complex fading coefficients $\{\mathbf{H}_{ki},\forall
k,i\}$ and path loss $\{L_{ki},\forall k,i\}$.}
Observe that the average SER
of all the schemes decreases as the receive $E_s/N_0$ increases, and
there is significant performance gain of the proposed schemes
compared to all baselines, even for low complexity PIAID with
SDR-SID (Algorithm \ref{alg:sdr-sic}). The performance gain is
contributed by the user selection of the PIA stage that moves the ID
processing out of the unfavorable interference profile as shown in
Fig. \ref{fig:ser_ic}. Furthermore, it can also be observed that PIAID
with SDR-SID (Algorithm \ref{alg:sdr-sic}) has similar performance
as PIAID (Algorithm \ref{alg:stage-I}). { Finally, Fig. \ref{fig:5users} shows that the proposed PIA algorithm with condition \eqref{eq:ia_feasibility} has similar performance as the solution obtained by brute-force exhaustive search (without \eqref{eq:ia_feasibility}).}

\subsection{Cumulative Distribution Function (CDF) of the SER}
\begin{figure}
\centering
\includegraphics[width = 14cm]{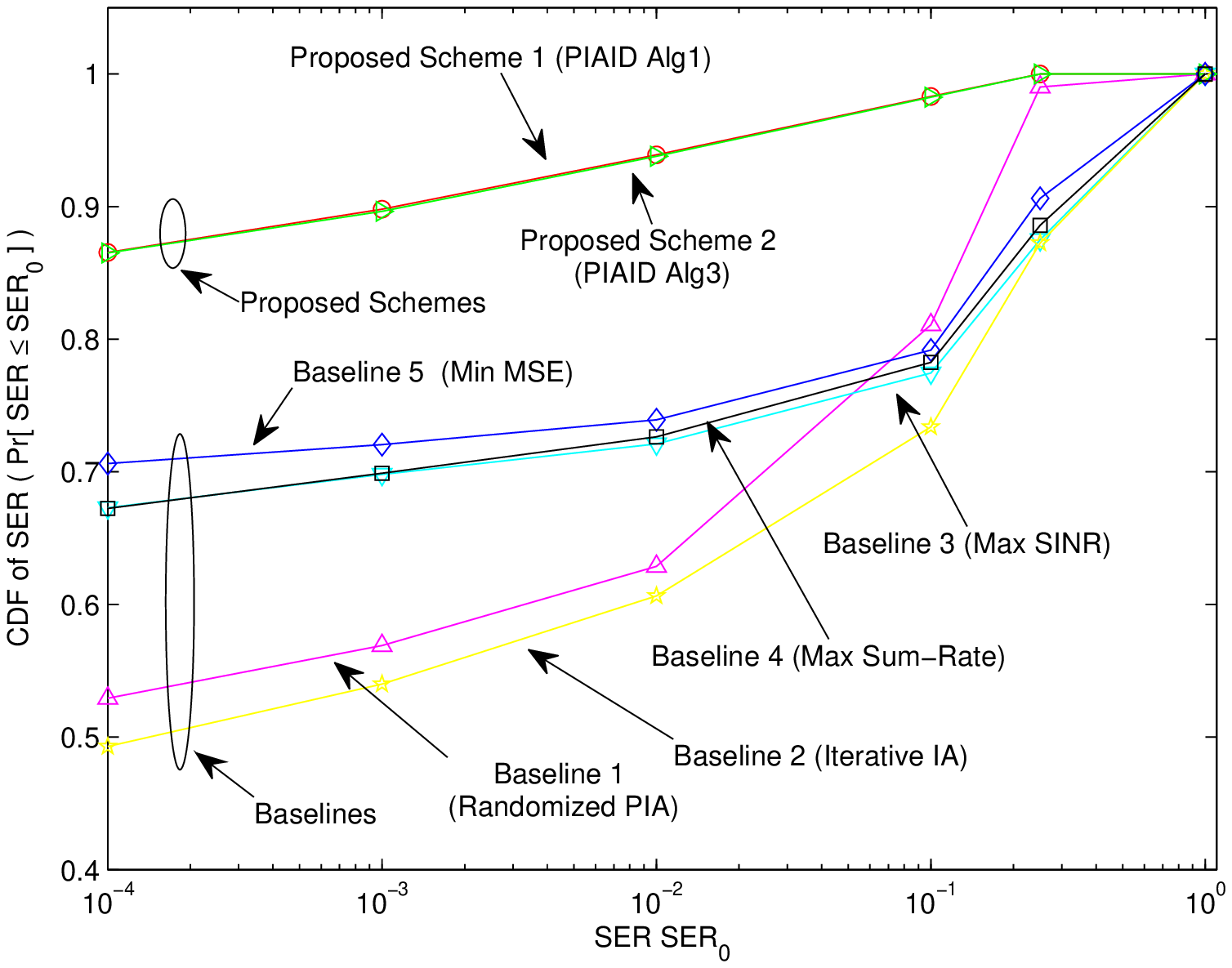}
\caption{ Cumulative Distribution Function (CDF) of the SER per data
stream with receive $E_s/N_0=25$dB. The setup is given by $K = 5$
(number of users), $\{M=3,N=2\}$ (number of transmit and receive
antennas), $D = 1$ (number of data stream), and $\alpha=3$ (number
of aligned users for feasible interference alignment).
}\label{fig:cdf_5users}
\end{figure}

\begin{figure}
\centering
\includegraphics[width = 14cm]{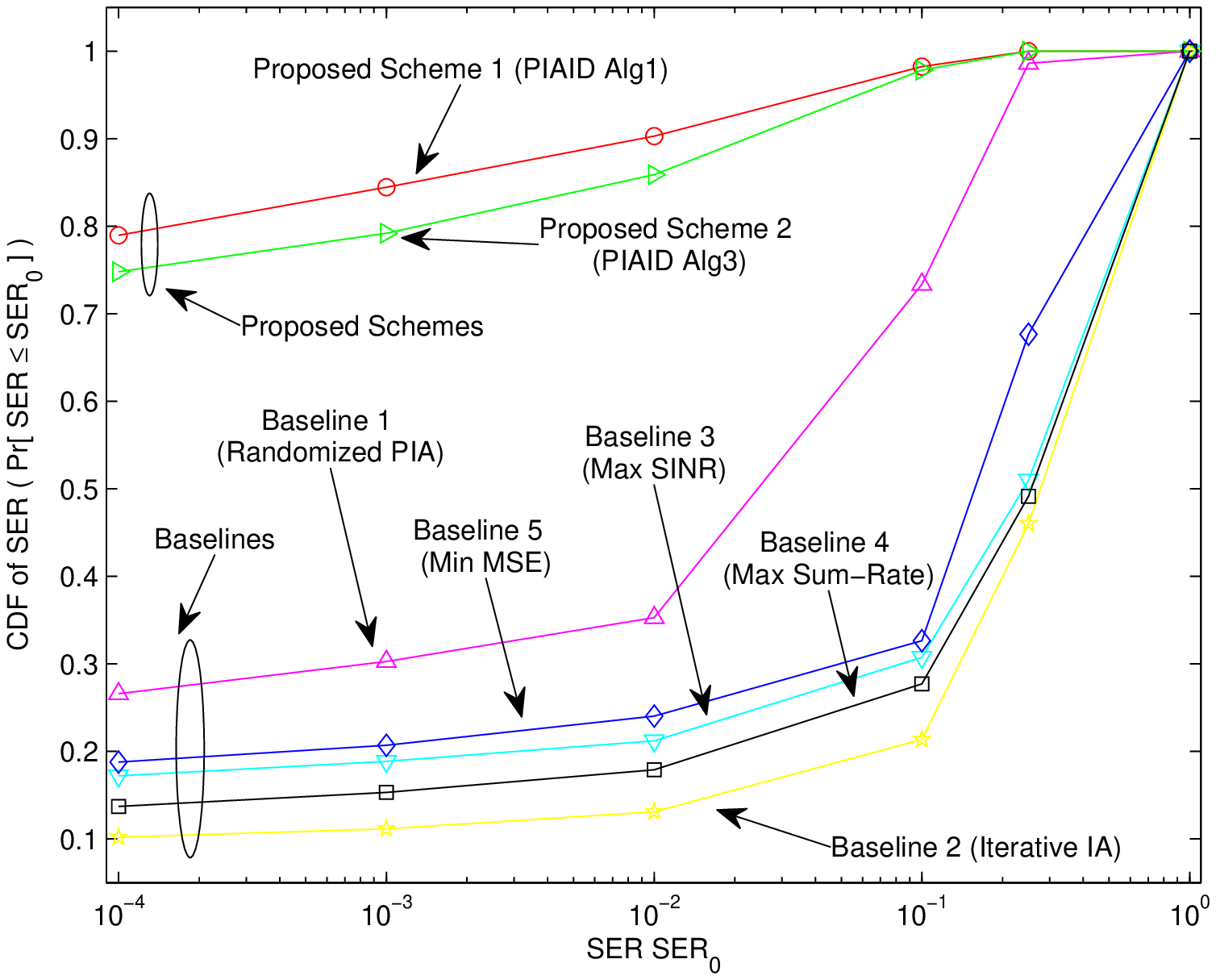}
\caption{ Cumulative Distribution Function (CDF) of the SER per data
stream with receive $E_s/N_0=25$dB. The setup is given by $K = 6$
(number of users), $\{M=3,N=2\}$ (number of transmit and receive
antennas), $D = 1$ (number of data stream), and $\alpha=3$ (number
of aligned users for feasible interference alignment).
}\label{fig:cdf_6users}
\end{figure}

Fig. \ref{fig:cdf_5users} and Fig. \ref{fig:cdf_6users} illustrate
the Cumulative Distribution Function (CDF) of the SER per data
stream with receive $E_s/N_0=25$dB for $K=5$ and $K=6$ respectively,
where the randomness of SER is induced by $\{\mathbf{H}_{ki},\forall
k,i\}$ and $\{L_{ki},\forall k,i\}$. { The number of transmit and receive antennas is given by
$\{M=3,N=2\}$. The number of aligned users for the feasible
interference alignment is $\alpha=3$. The CDF performance is evaluated with $10^7$ realizations of noise, complex fading coefficients $\{\mathbf{H}_{ki},\forall
k,i\}$ and path loss $\{L_{ki},\forall k,i\}$.}
It can also be verified that
the proposed scheme achieves not only a smaller average SER but also
a smaller SER percentile compared with the baselines.

\section{Conclusions}\label{sec:con}
In this paper, we propose a low complexity and novel Partial
Interference Alignment (PIA) and Interference Detection (PIAID)
scheme for $K$-user quasi-static MIMO interference channels with
general irrational channel coefficients. Based on the path loss
information, the proposed PIAID scheme dynamically selects the IA
interferers at each receiver such that it moves the ID processing
out of the unfavorable interference profiles. We derive the average
SER by taking into account the non-Guassian residual interference,
and obtain a low complexity user set selection
algorithm for the PIAID scheme, { which minimizes the asymptotically tight bound for the average
end-to-end SER performance.} The SER
performance of the proposed PIAID scheme is shown to have
significant gain compared with various conventional baseline
solutions.

\appendices

\section{Proof of Lemma \ref{lem:lp}}\label{app:lem:lp}
By introducing the slack variables ${s_{ki}}$, the LP relaxation
problem of (\ref{eq:sub_opt}) becomes
\begin{equation}
\{e_{ki}^*\}=\left\{
\begin{array}{cl}
\argmax_{e_{ki}} & \sum_{k,i}c_{ki}e_{ki} \\
\text{s.t} & \sum_{i}e_{ki} = \alpha, \forall k\\
&\sum_{k}e_{ki} = \alpha, \forall i\\
&e_{ki}+s_{ki} = 1, \forall k,i \\
&e_{ki}\geq0,s_{ki}\geq0, \forall k,i\end{array}\right.
\end{equation}
{ and it is equivalent to the following matrix form}
\begin{equation}\label{eq:LP}
\{\mathbf{e}^*\}=\left\{
\begin{array}{cl}
\argmax_{\mathbf{e}} & \mathbf{c}^T\mathbf{e}  \\
\text{s.t} & \mathbf{Ae}=\mathbf{b} \\
&\mathbf{e}\succeq0
\end{array}\right.
\end{equation}
where
$\mathbf{e}=[e_{11},\cdots,e_{1K},e_{21},\cdots,e_{KK},s_{11},\cdots,s_{1K},s_{21},\cdots,s_{KK}]^T$ { is formed by the optimization variables},
$\mathbf{b}=[\underbrace{\alpha,\cdots,\alpha}_{2K},\underbrace{1,\cdots,1}_{K^2}]^T$,
$\mathbf{c}=[c_{11},\cdots,c_{1K},
c_{21},\cdots,c_{KK},\underbrace{0,\cdots,0}_{K^2}]^T$ { is formed by the path costs},
and the matrix $\mathbf{A}$ is given by
\begin{equation}
\mathbf{A}=\left[
\begin{array}{c|c}
\mathbf{B} & \mathbf{0}_{2K\times K^2}\\
\hline
\mathbf{I}_{K^2} & \mathbf{I}_{K^2}\\
\end{array}\right]
=\left[
\begin{array}{c|c}
\begin{array}{cccc}
\mathbf{1}_{1\times K} & \mathbf{0}_{1\times K}  & \cdots & \mathbf{0}_{1\times K} \\
\mathbf{0}_{1\times K} & \mathbf{1}_{1\times K}  & \cdots & \mathbf{0}_{1\times K} \\
\vdots & \vdots & \vdots & \vdots \\
\mathbf{0}_{1\times K} & \mathbf{0}_{1\times K} & \cdots &
\mathbf{1}_{1\times K}\\
\mathbf{I}_{K} & \mathbf{I}_{K} & \cdots & \mathbf{I}_{K}
\end{array}
& \mathbf{0}_{2K\times K^2}\\
\hline
\mathbf{I}_{K^2} & \mathbf{I}_{K^2}\\
\end{array}\right].
\end{equation}
where $\mathbf{0}_{n\times
m}$ denotes an $n\times m$ matrix of zeros and $\mathbf{I}_{n}$
denotes an $n\times n$ identity matrix.

Note that the feasible set for the LP (\ref{eq:LP}) is given by the
polytope
$\mathcal{R}(\mathbf{A})=\{\mathbf{Ae}=\mathbf{b},\mathbf{e}\succeq0\}$.
If all the vertices of $\mathcal{R}(\mathbf{A})$ are integers, there
exists an optimal solution such that all the optimization variables
are integers, and hence the LP (\ref{eq:LP}) will always lead to an
integer optimum when solved by the well known simplex algorithm.
Since $e_{ki}\leq 1,\forall k,i$, if $e_{ki}$ is an integer, we have
$e_{ki}\in\{0,1\}$. As a result, it is equivalent to proving that
all the vertices of the polytope $\mathcal{R}(\mathbf{A})$ are
integers. We first define {\em Totally Unimodular} (TUM) as follows.
\begin{Def}[Definition of Totally Unimodular]
An integer matrix $\mathbf{Z}$ is totally unimodular if the
determinant of each square submatrix of $\mathbf{Z}$ is equal to
0,1, or -1. ~\hfill\IEEEQED
\end{Def}

It has been shown in \cite{comb:1998} that if $\mathbf{A}$ is TUM,
then all the vertices of $\mathcal{R}(\mathbf{A})$ are integers.
Therefore, the proof reduces to proving that $\mathbf{A}$ is TUM.
Note that the matrix $\mathbf{B}$ satisfies the following
conditions:
\begin{itemize}
\item $\mathbf{B}$ is a $\{0,1\}$ matrix with no more than two nonzero
elements in each column.
\item Each column
contains two nonzero elements that have the same sign, where one
element is in a row contained in the subset
$\Omega_1=\{1,\cdots,K\}$ and the other element is in a row
contained in the subset $\Omega_2=\{K+1,\cdots,2K\}$.
\end{itemize}
Therefore, $\mathbf{B}$ is TUM\cite{comb:1988}. It is easy to verify
that the matrix $\mathbf{C}=[\mathbf{B}^T,\mathbf{I}]^T$ is also
TUM. Let $\mathbf{D}$ be a square, nonsingular submatrix of
$\mathbf{A}$. The rows of $\mathbf{D}$ can be permuted so that it
can be written as
\begin{equation}
\mathbf{D}=\left[
\begin{array}{c|c}
\mathbf{E} & \mathbf{0}\\
\hline \mathbf{F} & \mathbf{I}
\end{array}\right],
\end{equation}
where $\mathbf{E}$ is a square submatrix of $\mathbf{C}$, and
possibly with its rows permuted. Therefore,
$\det(\mathbf{D})=\det(\mathbf{B})=\pm1$, which completes the proof.

\section{Proof of Theorem \ref{thm:ser}}\label{app:thm:ser}
\subsection{Upper Bound of the Average SER Performance}

In this subsection, we shall obtain an upper bound of the average
SER $\overline{\text{SER}}_k^l(\mathcal{A})$. { Specifically, we shall first
obtain an upper bound of the SER under a given
channel realization, i.e., an upper bound of
$\text{SER}_k^l(\mathcal{A},\mathcal{H})$.
\subsubsection{Upper bound of $\text{SER}_k^l(\mathcal{A},\mathcal{H})$}
Let
$\mathcal{H}_k^l=\{H_{ki}^{ld},\forall i,d\}$ denotes the set of equivalent channel gains after applying the equalizer $\mathbf{u}_k^l$ when decoding the desired signal $x_k^l$ at the receiver $k$,
$\delta_k^l=I_k^l-\hat{I}_k^l$ denotes the difference between the real and estimated aggregate strong interference, and
$\zeta_k^l=\sum_{i\in\mathcal{O}_k,n}\sqrt{P_{i}L_{ki}}H_{ki}^{ld}x_i^d$ denotes the aggregate weak residual interference. Specifically, we start to consider the two stage decoding separately.}
\begin{itemize}{
\item{\bf Stage II decoding:} In stage II decoding, $\widetilde{y}_k^l$ in (\ref{eq:y_k^l_2nd}) is
given by:
\begin{equation}
\widetilde{y}_k^l=y_k^l-\hat{I}_k^l=\sqrt{P_{k}L_{kk}}H_{kk}^{ll}x_k^l+\delta_k^l+\zeta_k^l
+(\mathbf{u}_k^l)^{\dag}\mathbf{z}_k,
\end{equation} }
Given $\mathcal{H}_k^l$ and suppose
$\Re\{x_k^l\}=\frac{\sqrt{2}}{2}$, the error rate of decoding
$\Re\{x_k^l\}$ is given by \cite{proakis:2001}:
\begin{equation}\label{eq:pr_err}
\begin{array}{l}
\Pr\Big\{\Re\{\hat{x}_k^l\}\neq\frac{\sqrt{2}}{2}|\Re\{x_k^l\}=\frac{\sqrt{2}}{2},I_k^l,\hat{I}_k^l,\mathcal{H}_k^l\Big\}
=\Pr\Big\{\Re\big\{\frac{\widetilde{y}_k^l}{\sqrt{P_{k}L_{kk}}H_{kk}^{ll}}\big\}
\leq 0|\Re\{x_k^l\}=\frac{\sqrt{2}}{2},I_k^l,\hat{I}_k^l,\mathcal{H}_k^l\Big\}\\
=Q\left(
\big(\frac{\sqrt{2}}{2}+\Re\{\frac{\delta_k^l+\zeta_k^l}{\sqrt{P_{k}L_{kk}}H_{kk}^{ll}}\}\big)
|\sqrt{P_{k}L_{kk}}H_{kk}^{ll}|\right)=Q\left(
\frac{\sqrt{2}}{2}|\sqrt{P_{k}L_{kk}}H_{kk}^{ll}|+\Re\{e^{-j\theta}(\delta_k^l+\zeta_k^l)\}\
\right),
\end{array}
\end{equation}
where $Q(x)=\frac{1}{\sqrt{2\pi}}\int_{x}^{\infty}e^{-x^2/2}\text{d}x$
denotes the $Q$-function, and
$e^{-j\theta}=\frac{\sqrt{P_{k}L_{kk}}H_{kk}^{ll}}{|\sqrt{P_{k}L_{kk}}H_{kk}^{ll}|}$.
Similarly, we have
\begin{equation}
\begin{array}{l}
\Pr\{\Re\{\hat{x}_k^l\}\neq-\frac{\sqrt{2}}{2}|\Re\{x_k^l\}=-\frac{\sqrt{2}}{2},I_k^l,\hat{I}_k^l,\mathcal{H}_k^l\}=Q\left(
\frac{\sqrt{2}}{2}|\sqrt{P_{k}L_{kk}}H_{kk}^{ll}|-\Re\{e^{-j\theta}(\delta_k^l+\zeta_k^l)\}\right),\\
\Pr\{\Im\{\hat{x}_k^l\}\neq\frac{\sqrt{2}}{2}|\Im\{x_k^l\}=\frac{\sqrt{2}}{2},I_k^l,\hat{I}_k^l,\mathcal{H}_k^l\}=Q\left(
\frac{\sqrt{2}}{2}|\sqrt{P_{k}L_{kk}}H_{kk}^{ll}|+\Im\{e^{-j\theta}(\delta_k^l+\zeta_k^l)\}\right),\\
\Pr\{\Im\{\hat{x}_k^l\}\neq-\frac{\sqrt{2}}{2}|\Im\{x_k^l\}=-\frac{\sqrt{2}}{2},I_k^l,\hat{I}_k^l,\mathcal{H}_k^l\}=Q\left(
\frac{\sqrt{2}}{2}|\sqrt{P_{k}L_{kk}}H_{kk}^{ll}|-\Im\{e^{-j\theta}(\delta_k^l+\zeta_k^l)\}
\right).
\end{array}
\end{equation}
{ Then the error rate of decoding $x_k^l$ is given by:
\begin{equation}
\begin{array}{ll}\label{eq:pr_err_symbol}
\quad\Pr\{\hat{x}_k^l\neq x_k^l|x_k^l,I_k^l,\hat{I}_k^l,\mathcal{H}_k^l\}\\
\leq\Pr\{\Re\{\hat{x}_k^l\}\neq\Re\{x_k^l\}|\Re\{x_k^l\},I_k^l,\hat{I}_k^l,\mathcal{H}_k^l\}+
\Pr\{\Im\{\hat{x}_k^l\}\neq\Im\{x_k^l\}|\Im\{x_k^l\},I_k^l,\hat{I}_k^l,\mathcal{H}_k^l\}\\
\leq2Q\left(
\frac{\sqrt{2}}{2}|\sqrt{P_{k}L_{kk}}H_{kk}^{ll}|-|\delta_k^l|-|\zeta_k^l|\right).
\end{array}
\end{equation}
\item{\bf Stage I decoding:} In the stage I decoding, $y_k^l$ in (\ref{eq:y_k^l_1st}) is given
by:
\begin{equation}
y_k^l=I_k^l+\sqrt{P_{k}L_{kk}}H_{kk}^{ll}x_k^l+\zeta_k^l+
(\mathbf{u}_k^l)^{\dag}\mathbf{z}_k.
\end{equation}
Note that when the detected aggregate strong interference $\hat{I}_k^l\neq I_k^l$,
an upper bound of the probability that the estimated aggregate strong interference is $\hat{I}_k^l$ is given by
\begin{equation}
\Pr\{\hat{I}_k^l|x_k^l,I_k^l,\mathcal{H}_k^l\}\leq
\Pr\{\hat{I}_k^l\neq I_k^l|x_k^l,I_k^l,\mathcal{H}_k^l\}\leq 2Q\left(
|\delta_k^l|/2-|\sqrt{P_{k}L_{kk}}H_{kk}^{ll}|-|\zeta_k^l|\right),
\end{equation}
where $\Pr\{\hat{I}_k^l\neq I_k^l|x_k^l,I_k^l,\mathcal{H}_k^l\}$ is the conditional probability that $\hat{I}_k^l\neq
I_k^l$ when $I_k^l$ is transmitted. Therefore, we can follow the
same way as (\ref{eq:pr_err}) and (\ref{eq:pr_err_symbol}), which obtain the conditional probability
that $\hat{x}_k^l\neq x_k^l$ when $x_k^l$ is transmitted. When the detected aggregate strong interference $\hat{I}_k^l= I_k^l$, obviously we have following
expression at interference limited regime
\begin{equation}
\Pr\{\hat{I}_k^l|x_k^l,I_k^l,\mathcal{H}_k^l\}=
\Pr\{I_k^l|x_k^l,I_k^l,\mathcal{H}_k^l\}\leq
2Q\left(-|\sqrt{P_{k}L_{kk}}H_{kk}^{ll}|-|\zeta_k^l|\right)\approx 2.
\end{equation}
As a result, the probability that the estimated aggregate strong interference is $\hat{I}_k^l$ in the stage I decoding is given by
\begin{equation}\label{eq:pr_err_symbol_stage_II}
\Pr\{\hat{I}_k^l|x_k^l,I_k^l,\mathcal{H}_k^l\}\leq 2Q\left(
|\delta_k^l|/2-|\sqrt{P_{k}L_{kk}}H_{kk}^{ll}|-|\zeta_k^l|\right).
\end{equation}}
\end{itemize}
\vspace{-20pt}

From the upper bound expressions (\ref{eq:pr_err_symbol_stage_II})
and (\ref{eq:pr_err_symbol}) in the stage I
and stage II decoding, respectively, an upper bound of the SER under a given
channel realization $\text{SER}_k^l(\mathcal{A},\mathcal{H})$ is
given by
\begin{equation}\label{eq:ser_channel}
\begin{array}{lll}
&\text{SER}_k^l(\mathcal{A},\mathcal{H})\triangleq\sum_{x_k^l\in\mathcal{S}}\Pr\{x_k^l\}\Pr\{\hat{x}_k^l\neq
x_k^l|x_k^l,\mathcal{H}_k^l\}\\
=&\sum_{x_k^l\in\mathcal{S}}\Pr\{x_k^l\}\left(\sum_{I_k^l\in\mathcal{I}_k^l}\Pr\{I_k^l\}\Pr\{\hat{x}_k^l\neq
x_k^l|x_k^l,I_k^l,\mathcal{H}_k^l\}\right)\\
=&\sum_{x_k^l\in\mathcal{S}}\Pr\{x_k^l\}\left(\sum_{I_k^l\in\mathcal{I}_k^l}\sum_{\hat{I}_k^l\in\mathcal{I}_k^l}
\Pr\{I_k^l\}\Pr\{\hat{I}_k^l|x_k^l,I_k^l,\mathcal{H}_k^l\}\Pr\{\hat{x}_k^l\neq
x_k^l|x_k^l,I_k^l,\hat{I}_k^l,\mathcal{H}_k^l\}\right)\\
\leq&\frac{1}{|\mathcal{I}_k|}\sum_{x_k^l\in\mathcal{S}}\sum_{I_k^l\in\mathcal{I}_k^l}
\sum_{\hat{I}_k^l\in\mathcal{I}_k^l}Q\left(
|\delta_k^l|/2-|\sqrt{P_{k}L_{kk}}H_{kk}^{ll}|-|\zeta_k^l|\right)\cdot\\
&Q\left(
\frac{\sqrt{2}}{2}|\sqrt{P_{k}L_{kk}}H_{kk}^{ll}|-|\delta_k^l|-|\zeta_k^l|\right).
\end{array}
\end{equation}

{ \subsubsection{Upper bound of $\text{SER}_k^l(\mathcal{A})$}
From (\ref{eq:ser_channel}), the average SER is given by }
\begin{equation}\label{eq:ser_upper_org}
\begin{array}{lll}
&\overline{\text{SER}}_k^l(\mathcal{A})\triangleq
\mathbb{E}_{\mathcal{H}}\left[\text{SER}_k^l(\mathcal{A},\mathcal{H})\right]\\
\leq&\frac{1}{|\mathcal{I}_k|}\mathbb{E}_{\mathcal{H}_k^l}\Big[\sum_{I_k^l\in\mathcal{I}_k^l}\sum_{\hat{I}_k^l\in\mathcal{I}_k^l}
\mathbf{1}
\big(|\delta_k^l|/2-|\sqrt{P_{k}L_{kk}}H_{kk}^{ll}|-|\zeta_k^l|<0\big)\cdot\\
&\quad\quad\quad\quad\mathbf{1}
\big(\frac{\sqrt{2}}{2}|\sqrt{P_{k}L_{kk}}H_{kk}^{ll}|-|\delta_k^l|-|\zeta_k^l|<0\big)\Big] \\
=&\frac{1}{|\mathcal{I}_k|}\sum_{I_k^l\in\mathcal{I}_k^l}\sum_{\hat{I}_k^l\in\mathcal{I}_k^l}
\Pr\Big\{\frac{\sqrt{2}}{2}|\sqrt{P_{k}L_{kk}}H_{kk}^{ll}|-|\zeta_k^l|\leq|\delta_k^l|\leq2(|\sqrt{P_{k}L_{kk}}H_{kk}^{ll}|-|\zeta_k^l|)\Big\}.
\end{array}
\end{equation}

{ Therefore, from the above equation (\ref{eq:ser_upper_org}), we
shall discuss the following two cases given by: successful stage I decoding $\hat{I}_k^l=I_k^l$
(i.e., $\delta_k^l=0$) and unsuccessful stage I decoding $\hat{I}_k^l\neq I_k^l$ (i.e.,
$\delta_k^l\neq0$).}
\begin{itemize}{
\item {\bf Successful stage I decoding $\hat{I}_k^l=I_k^l$
($\delta_k^l=0$):} Note that
$H_{ki}^{ld}=(\mathbf{u}_k^l)^{\dag}\mathbf{H}_{ki}\mathbf{v}_i^d$,
and the aggregate weak residual interference $\zeta_k^l$ is given by}
\begin{equation}
\begin{array}{rl}
\zeta_k^l=&\sum_{i\in\mathcal{O}_k,d}\sqrt{P_{i}L_{ki}}H_{ki}^{ld}x_i^d=
\sum_{i\in\mathcal{O}_k,d}\sqrt{P_{i}L_{ki}}(\mathbf{u}_k^l)^{\dag}\mathbf{H}_{ki}\mathbf{v}_i^dx_i^d\\
=&\sum_{i\in\mathcal{O}_k}\sqrt{P_{i}L_{ki}}\sum_{m,n}
[(\mathbf{u}_k^l)^{\dag}]_{(1,n)}[\mathbf{H}_{ki}]_{(n,m)}\sum_{d}[\mathbf{v}_i^dx_i^d]_{(m,1)}.
\end{array}
\end{equation}

Since $\{\mathbf{u}_k^l,\mathbf{v}_i^d,\forall
i\in\mathcal{O}_k,\forall d\}$ are determined by the channel gains
from the aligned links given by $\{\mathbf{H}_{ki},\forall
i\in\mathcal{A}_k,\forall
k\}$\cite{Feasibility:MIMO:2009,Tresch:2009}, they are independent
of the random variables $\{\mathbf{H}_{ki},\forall
i\in\mathcal{O}_k\}$. From Assumption \ref{ass:channel_model}, $[\mathbf{H}_{ki}]_{(n,m)}\sim\mathcal{CN}(0,1)$, and hence given
$\{\mathbf{u}_k^l,\mathbf{v}_i^dx_i^d,\forall
i\in\mathcal{O}_k,\forall d\}$, $\zeta_k^l$ is a complex Gaussian
variable with zero mean and variance
\begin{equation}
\begin{array}{rl}
\sigma^2=&\sum_{i\in\mathcal{O}_k}P_{i}L_{ki}\sum_{m,n}\big|
[(\mathbf{u}_k^l)^{\dag}]_{(1,n)}\big|^2\left(\big|\sum_{d}[\mathbf{v}_i^dx_i^d]_{(m,1)}\big|\right)^2\\
=&\sum_{i\in\mathcal{O}_k}P_{i}L_{ki}\sum_{m}\left(\big|\sum_{d}[\mathbf{v}_i^dx_i^d]_{(m,1)}\big|\right)^2
\leq D\sum_{i\in\mathcal{O}_k}P_{i}L_{ki}\sum_{m}\left(\sum_{d}\big|[\mathbf{v}_i^dx_i^d]_{(m,1)}\big|^2\right)\\
=&D^2\sum_{i\in\mathcal{O}_k}P_{i}L_{ki},
\end{array}
\end{equation}
because of $||\mathbf{u}_k^l||=||\mathbf{v}_i^dx_i^d||=1$.
From $\zeta_k^l$ is a complex Gaussian
variable with zero mean and variance $\sigma^2$, and the probability density function of the random variable
$Z=|\frac{\zeta_k^l}{\sigma H_{kk}^{ll}}|^2$ is given by $f_{Z}(z) =
\frac{1}{(z+1)^2}$. { Therefore, when $\delta_k^l=0$ the conditional probability is given by:}
\begin{equation}
\begin{array}{lll}
&&\Pr\Big\{\frac{\sqrt{2}}{2}|\sqrt{P_{k}L_{kk}}H_{kk}^{ll}|-|\zeta_k^l|\leq0\big|\mathbf{u}_k^l,\mathbf{v}_i^dx_i^d,\forall
i\in\mathcal{O}_k,\forall d\Big\}=
\Pr\{\frac{P_{k}L_{kk}}{2\sigma^2}\leq
Z\}\\
&=&1/(\frac{P_{k}L_{kk}}{2\sigma^2}+1)\leq1/(\frac{P_{k}L_{kk}}{4D\sum_{i\in\mathcal{O}_k}P_{i}L_{ki}}+1)=\ThetaL\left(
\sum_{i\in\mathcal{O}_k}\frac{P_{i}L_{ki}}{P_{k}L_{kk}}\right).
\end{array}
\end{equation}
{ In turn, the probability of the event $\frac{\sqrt{2}}{2}|\sqrt{P_{k}L_{kk}}H_{kk}^{ll}|\leq|\zeta_k^l|$ is given by}
\begin{equation}\label{eq:ser_upper_0}
\begin{array}{lll}
\Pr\Big\{\frac{\sqrt{2}}{2}|\sqrt{P_{k}L_{kk}}H_{kk}^{ll}|-|\zeta_k^l|\leq0\Big\}
\leq
\ThetaL\left(
\sum_{i\in\mathcal{O}_k}\frac{P_{i}L_{ki}}{P_{k}L_{kk}}\right).
\end{array}
\end{equation}

\item {  {\bf Unsuccessful stage I decoding $\hat{I}_k^l\neq I_k^l$ ($\delta_k^l\neq0$):} Given
$\hat{I}_k^l=\sum_{i\in\mathcal{Q}_k;d}\sqrt{P_{i}L_{ki}}H_{ki}^{ld}\hat{x}_i^d$, where $\hat{x}_i^d\in\mathcal{S}$,
$\delta_k^l$ is given by:}
\begin{equation}
\begin{array}{rl}
\delta_k^l=&I_k^l-\hat{I}_k^l=\sum_{i\in\mathcal{Q}_k;d}\sqrt{P_{i}L_{ki}}
(\mathbf{u}_k^l)^{\dag}\mathbf{H}_{ki}\mathbf{v}_i^d(x_i^d-\hat{x}_i^d)\\
=&\sum_{i\in\mathcal{Q}_k}\sqrt{P_{i}L_{ki}}\sum_{m,n}
[(\mathbf{u}_k^l)^{\dag}]_{(1,n)}[\mathbf{H}_{ki}]_{(n,m)}\sum_{d}[\mathbf{v}_i^d\psi_i^d]_{(m,1)},
\end{array}
\end{equation}
where $\psi_i^d=x_i^d-\hat{x}_i^d\in\{0,\pm\sqrt{2}\pm
j\sqrt{2},\pm\sqrt{2},\pm j\sqrt{2}\}$. Given
$\{\mathbf{u}_k^l,\mathbf{v}_i^d\psi_i^d,\forall
i\in\mathcal{Q}_k,\forall d\}$, $\delta_k^l$ is a complex Gaussian
variable with zero mean and variance given by
\begin{equation}
\begin{array}{rl}
\sigma^2=&\sum_{i\in\mathcal{Q}_k}P_{i}L_{ki}\sum_{m,n}\big|
[(\mathbf{u}_k^l)^{\dag}]_{(1,n)}\big|^2\big|\sum_{d}[\mathbf{v}_i^d\psi_i^d]_{(m,1)}\big|^2\\
=&
\sum_{i\in\mathcal{Q}_k}P_{i}L_{ki}\sum_{m}\big|\sum_{d}[\mathbf{v}_i^d\psi_i^d]_{(m,1)}\big|^2
\geq
P_{i^*}L_{ki^*}\sum_{i\in\mathcal{Q}_k}\sum_{m}\big|\sum_{d}[\mathbf{v}_i^d\psi_i^d]_{(m,1)}\big|^2,
\end{array}
\end{equation}
where $i^*=\argmin_{i\in\Xi} P_{i}L_{ki}$, and
$\Xi=\{i:\sum_{d}|\psi_i^d|\neq0\}$, since $||\mathbf{u}_k^l||=1$
and $||\mathbf{v}_i^d\psi_i^d||= |\psi_i^d|$. { Therefore, when
$\delta_k^l\neq 0$, the conditional probability is given by}
\begin{equation}
\begin{array}{ll}
&\Pr\Big\{\frac{\sqrt{2}}{2}|\sqrt{P_{k}L_{kk}}H_{kk}^{ll}|-|\zeta_k^l|\leq|\delta_k^l|
\leq2(|\sqrt{P_{k}L_{kk}}H_{kk}^{ll}|-|\zeta_k^l|)\big|\mathbf{u}_k^l,\mathbf{v}_i^d\psi_i^d,\forall
i\in\mathcal{Q}_k,\forall d\Big\}\\
\leq&\Pr\Big\{0\leq|\delta_k^l|\leq2|\sqrt{P_{k}L_{kk}}H_{kk}^{ll}|\Big\}=
\Pr\Big\{0 \leq|\frac{\delta_k^l}{\sigma H_{kk}^{ll}}|^2
\leq\frac{4P_{k}L_{kk}}{\sigma^2})\Big\}\\
=&\frac{4P_{k}L_{kk}/\sigma^2}{4P_{k}L_{kk}/\sigma^2+1}<\frac{4P_{k}L_{kk}}{\sigma^2}.
\end{array}
\end{equation}
{ In turn, the unconditional probability is given by:}
\begin{equation}
\begin{array}{ll}\label{eq:ser_upper_neq0}
&\Pr\Big\{\frac{\sqrt{2}}{2}|\sqrt{P_{k}L_{kk}}H_{kk}^{ll}|-|\zeta_k^l|\leq|\delta_k^l|
\leq2(|\sqrt{P_{k}L_{kk}}H_{kk}^{ll}|-|\zeta_k^l|)\Big\}\\
\leq&\mathbb{E}_{\mathbf{u}_k^l,\mathbf{v}_i^d\psi_i^d}\left[\frac{4P_{k}L_{kk}}{\sigma^2}\right]=C\frac{P_{k}L_{kk}}{P_{i^*}L_{ki^*}},
\end{array}
\end{equation}
where $C=\mathbb{E}_{\mathbf{u}_k^l,\mathbf{v}_i^d\psi_i^d}
\left[4/(\sum_{i\in\mathcal{Q}_k}\sum_{m}|\sum_{d}[\mathbf{v}_i^d\psi_i^d]_{(m,1)}^2)
\right]$ is a positive constant.
\end{itemize}

Finally, substitute (\ref{eq:ser_upper_0}) and
(\ref{eq:ser_upper_neq0}) into (\ref{eq:ser_upper_org}), we have
\begin{equation}
\begin{array}{l}
\overline{\text{SER}}_k^l(\mathcal{A})\leq\ThetaL\left(
\sum_{i\in\mathcal{Q}_k}\frac{P_{k}L_{kk}}{P_{i}L_{ki}}+\sum_{i\in\mathcal{O}_k}\frac{P_{i}L_{ki}}{P_{k}L_{kk}}\right).
\end{array}
\end{equation}

\subsection{Lower Bound of the Average SER Performance}
{ In this subsection, we shall obtain a lower bound of the average SER
$\overline{\text{SER}}_k^l(\mathcal{A})$. Specifically, we shall prove $\overline{\text{SER}}_k^l(\mathcal{A})\geq\ThetaL(\frac{P_{i}L_{ki}}{P_{k}L_{kk}}),\forall i\in\mathcal{O}_k$ and
$\overline{\text{SER}}_k^l(\mathcal{A})\geq
\ThetaL(\frac{P_{k}L_{kk}}{P_{i}L_{ki}}),\forall
i\in\mathcal{Q}_k$, respectively.}

\begin{itemize}
\item { \bf Proof of
$\overline{\text{SER}}_k^l(\mathcal{A})\geq\ThetaL(\frac{P_{i}L_{ki}}{P_{k}L_{kk}}),\forall i\in\mathcal{O}_k$:} Suppose receiver $k$ has perfect
knowledge of all the data streams except $x_i^d,\forall
i\in\mathcal{O}_k,\forall d$. After cancelling the known data
streams in stage II decoding, $\widetilde{y}_k^l$ in (\ref{eq:y_k^l_2nd}) is
given by:
\begin{equation}
\widetilde{y}_k^l=\sqrt{P_{k}L_{kk}}H_{kk}^{ll}x_k^l+\sqrt{P_{i}L_{ki}}H_{ki}^{ld}x_i^d+
(\mathbf{u}_k^l)^{\dag}\mathbf{z}_k
\end{equation}

{ Suppose $\Re\{x_k^l\}=\frac{\sqrt{2}}{2}$, the error rate of
decoding $\Re\{x_k^l\}$ is given by (cf. (\ref{eq:pr_err})):}
\begin{equation}
\begin{array}{ll}
&\Pr\Big\{\Re\{\hat{x}_k^l\}\neq\Re\{x_k^l\}|\Re\{x_k^l\}=\frac{\sqrt{2}}{2}\Big\}\\
=&\sum_{x_i^d\in\mathcal{S}}\Pr\{x_i^d\}Q\left(
\frac{\sqrt{2}}{2}|\sqrt{P_{k}L_{kk}}H_{kk}^{ll}|+\Re\{e^{-j\theta}\sqrt{P_{i}L_{ki}}H_{ki}^{ld}x_i^d\}\right).
\end{array}
\end{equation}
where
$e^{j\theta}=\frac{\sqrt{P_{k}L_{kk}}H_{kk}^{ll}}{|\sqrt{P_{k}L_{kk}}H_{kk}^{ll}|}$.
Note that
$\Re\{e^{-j\theta}H_{ki}^{ld}x_i^d\}\in\mathcal{N}(0,\frac{1}{2}\sigma^2)$,
{ and hence the average error rate of decoding $\Re\{x_k^l\}$ is given by:}
\begin{equation}
\begin{array}{l}
\mathbb{E}\left[\Pr\{\Re\{\hat{x}_k^l\}\neq\Re\{x_k^l\}|\Re\{x_k^l\}=\frac{\sqrt{2}}{2}\}\right]=\Pr\Big\{
\frac{\sqrt{2}}{2}|\sqrt{P_{k}L_{kk}}H_{kk}^{ll}|+\Re\{e^{-j\theta}\sqrt{P_{i}L_{ki}}H_{ki}^{ld}x_i^d\}<0\Big\}\\
=\Pr\Big\{
|\sqrt{P_{k}L_{kk}}H_{kk}^{ll}|<\sqrt{P_{i}L_{ki}}\kappa\Big\}=\frac{1}{2}\left(1-\sqrt{\frac{1}{1+P_{i}L_{ki}/P_{k}L_{kk}}}\right)
=\frac{P_{i}L_{ki}}{4P_{k}L_{kk}}+O(\frac{P_{i}^2L_{ki}^2}{P_{k}^2L_{kk}^2})=\ThetaL(\frac{P_{i}L_{ki}}{P_{k}L_{kk}}),
\end{array}
\end{equation}
where $\kappa\in\mathcal{N}(0,\sigma^2)$ and
$|H_{kk}^{ll}|\in\text{Rayleigh}(\sigma^2)$. Similarly, we have
\begin{equation}
\begin{array}{ll}
&\mathbb{E}\left[\Pr\{\Re\{\hat{x}_k^l\}\neq\Re\{x_k^l\}|\Re\{x_k^l\}=-\frac{\sqrt{2}}{2}\}\right]\\
=&\Pr\{
\frac{\sqrt{2}}{2}|\sqrt{P_{k}L_{kk}}H_{kk}^{ll}|-\Re\{e^{-j\theta}\sqrt{P_{i}L_{ki}}H_{ki}^{ld}x_i^d\}<0\}
=\ThetaL(\frac{P_{i}L_{ki}}{P_{k}L_{kk}}),\\
&\mathbb{E}\left[\Pr\{\Im\{\hat{x}_k^l\}\neq\Im\{x_k^l\}|\Im\{x_k^l\}=\frac{\sqrt{2}}{2}\}\right]\\
=&\Pr\{
\frac{\sqrt{2}}{2}|\sqrt{P_{k}L_{kk}}H_{kk}^{ll}|+\Im\{e^{-j\theta}\sqrt{P_{i}L_{ki}}H_{ki}^{ld}x_i^d\}<0\}
=\ThetaL(\frac{P_{i}L_{ki}}{P_{k}L_{kk}}),\\
&\mathbb{E}\left[\Pr\{\Im\{\hat{x}_k^l\}\neq\Im\{x_k^l\}|\Im\{x_k^l\}=-\frac{\sqrt{2}}{2}\}\right]\\
=&\Pr\{
\frac{\sqrt{2}}{2}|\sqrt{P_{k}L_{kk}}H_{kk}^{ll}|-\Im\{e^{-j\theta}\sqrt{P_{i}L_{ki}}H_{ki}^{ld}x_i^d\}<0\}
=\ThetaL(\frac{P_{i}L_{ki}}{P_{k}L_{kk}}).
\end{array}
\end{equation}

Therefore, given that receiver $k$ has perfect
knowledge of all the data streams except $x_i^d,\forall
i\in\mathcal{O}_k,\forall d$, the average SER of decoding $x_k^l$ is given by
\begin{equation}
\overline{\text{SER}}_i^d\triangleq
\mathbb{E}\left[\sum\nolimits_{x_k^l\in\mathcal{S}}\Pr\{x_k^l\}\Pr\{\hat{x}_k^l\neq
x_k^l|x_k^l\}\right]
=\ThetaL\big(\frac{P_{i}L_{ki}}{P_{k}L_{kk}}\big),
\end{equation}
and obviously we have
$\overline{\text{SER}}_k^l(\mathcal{A})\geq\overline{\text{SER}}_i^d=
\ThetaL\left(\frac{P_{i}L_{ki}}{P_{k}L_{kk}}\right)$, $\forall
i\in\mathcal{O}_k$.

\item{\bf  Proof of
$\overline{\text{SER}}_k^l(\mathcal{A})\geq
\ThetaL(\frac{P_{k}L_{kk}}{P_{i}L_{ki}}),\forall
i\in\mathcal{Q}_k$:} Using similar arguments above and suppose
receiver $k$ has perfect knowledge of all the data streams except
$x_i^d,\forall i\in\mathcal{Q}_k,\forall d$. After cancelling the
known data streams in the stage I decoding, $y_k^l$ in (\ref{eq:y_k^l_1st}) is given
by:
\begin{equation}
y_k^l=\sqrt{P_{k}L_{kk}}H_{kk}^{ll}x_k^l+\sqrt{P_{i}L_{ki}}H_{ki}^{ld}x_i^d+
(\mathbf{u}_k^l)^{\dag}\mathbf{z}_k.
\end{equation}

Given
$\Re\{x_k^l\}=\frac{\sqrt{2}}{2},\Re\{x_i^d\}=\frac{\sqrt{2}}{2}$,
{ the error rate of decoding $\Re\{x_k^l\}$ is given by}
\begin{equation}
\begin{array}{ll}
&\Pr\Big\{\Re\{\hat{x}_k^l\}\neq\Re\{x_k^l\}|\Re\{x_k^l\},\Re\{x_i^d\}\Big\}\\
\geq&\Pr\Big\{\Re\{\hat{x}_i^d\}\neq\Re\{x_i^d\}|\Re\{x_k^l\},\Re\{x_i^d\}\Big\}
\Pr\Big\{\Re\{\hat{x}_k^l\}\neq\Re\{x_k^l\}|\Re\{x_k^l\},\Re\{x_i^d\},\Re\{\hat{x}_i^d\}\Big\}\\
=&Q\left(
\frac{\sqrt{2}}{2}|\sqrt{P_{i}L_{ki}}H_{ki}^{ld}|+\Re\{e^{-j\psi}\sqrt{P_{k}L_{kk}}H_{kk}^{ll}\sqrt{2}\}\right)\cdot\\
&\quad\quad\quad\quad Q\left(
\frac{\sqrt{2}}{2}|\sqrt{P_{k}L_{kk}}H_{kk}^{ll}|+\Re\{e^{-j\theta}\sqrt{P_{i}L_{ki}}H_{ki}^{ld}\sqrt{2}\}
\right).
\end{array}
\end{equation}
where
$e^{j\theta}=\frac{\sqrt{P_{k}L_{kk}}H_{kk}^{ll}}{|\sqrt{P_{k}L_{kk}}H_{kk}^{ll}|},
e^{j\psi}=\frac{\sqrt{P_{i}L_{ki}}H_{ki}^{ld}}{|\sqrt{P_{i}L_{ki}}H_{ki}^{ld}|},\vartheta=\theta-\psi$.
The probability density function of
$Z=|\frac{H_{kk}^{ll}}{H_{ki}^{ld}}|^2$ is given by $f_{Z}(z) =
\frac{1}{(z+1)^2}$, and $\vartheta$ follows the uniform distribution
between $0$ and $2\pi$. As a result, { the average error rate of decoding $\Re\{x_k^l\}$ is given by}
\begin{equation}
\begin{array}{ll}
&\mathbb{E}\left[\Pr\Big\{\Re\{\hat{x}_k^l\}\neq\Re\{x_k^l\}|\Re\{x_k^l\},\Re\{x_i^d\}\Big\}\right]\\
\geq&\Pr\left\{
\frac{1}{2\cos(\vartheta)}|\sqrt{P_{i}L_{ki}}H_{ki}^{ld}|<|\sqrt{P_{k}L_{kk}}H_{kk}^{ll}|<\cos(\vartheta)|\sqrt{P_{i}L_{ki}}H_{ki}^{ld}|\right\}\\
=&\int_{-\pi/4}^{\pi/4}\frac{(\cos^2(\vartheta)-1/(4\cos^2(\vartheta)))P_{k}L_{kk}/P_{i}L_{ki}}
{(1+P_{k}L_{kk}/(4P_{i}L_{ki}\cos^2(\vartheta)))(1+\cos^2(\vartheta)P_{k}L_{kk}/P_{i}L_{ki})}\text{d}\vartheta=\ThetaL(\frac{P_{k}L_{kk}}{P_{i}L_{ki}}).
\end{array}
\end{equation}
Furthermore, given that receiver $k$ has perfect knowledge of all the data streams except
$x_i^d,\forall i\in\mathcal{Q}_k,\forall d$, { the average error rate of decoding $x_k^l$ is given by}
\begin{equation}
\begin{array}{ll}
\overline{\text{SER}}_i^d\triangleq\mathbb{E}\left[\Pr\Big\{\hat{x}_k^l\neq x_k^l|x_k^l,x_i^d\Big\}\right]
\geq\ThetaL(\frac{P_{k}L_{kk}}{P_{i}L_{ki}}).
\end{array}
\end{equation}
Obviously we have
$\overline{\text{SER}}_k^l(\mathcal{A})\geq\overline{\text{SER}}_i^d\geq
\ThetaL(\frac{P_{k}L_{kk}}{P_{i}L_{ki}})$, $\forall
i\in\mathcal{Q}_k$.
\end{itemize}

Finally, from the results of upper and lower bound, we can conclude
that
\begin{equation}
\begin{array}{l}
\overline{\text{SER}}_k^l(\mathcal{A})\triangleq
\mathbb{E}_{\mathcal{H}}\left[\text{SER}_k^l(\mathcal{A},\mathcal{H})\right]\triangleq \ThetaL\left(\sum_{i\in\mathcal{Q}_k}\frac{P_{k}L_{kk}}{P_{i}L_{ki}}+
\sum_{i\in\mathcal{O}_k}\frac{P_{i}L_{ki}}{P_{k}L_{kk}}\right).
\end{array}
\end{equation}

\bibliographystyle{IEEEtran}
\bibliography{IEEEabrv,SER}

\begin{thebibliography}{10}
\providecommand{\url}[1]{#1}
\csname url@samestyle\endcsname
\providecommand{\newblock}{\relax}
\providecommand{\bibinfo}[2]{#2}
\providecommand{\BIBentrySTDinterwordspacing}{\spaceskip=0pt\relax}
\providecommand{\BIBentryALTinterwordstretchfactor}{4}
\providecommand{\BIBentryALTinterwordspacing}{\spaceskip=\fontdimen2\font plus
\BIBentryALTinterwordstretchfactor\fontdimen3\font minus
  \fontdimen4\font\relax}
\providecommand{\BIBforeignlanguage}[2]{{%
\expandafter\ifx\csname l@#1\endcsname\relax
\typeout{** WARNING: IEEEtran.bst: No hyphenation pattern has been}%
\typeout{** loaded for the language `#1'. Using the pattern for}%
\typeout{** the default language instead.}%
\else
\language=\csname l@#1\endcsname
\fi
#2}}
\providecommand{\BIBdecl}{\relax}
\BIBdecl

\bibitem{HK:gaussian_IC}
T.~S. Han and K.~Kobayashi, ``{A new achievable rate region for the
  interference channel},'' \emph{{IEEE} Trans. Inf. Theory}, vol.~27, pp.
  49--60, Jan. 1981.

\bibitem{IC:gaussian:capacity}
R.~H. Etkin, D.~N.~C. Tse, and H.~Wang, ``{Gaussian Interference Channel
  Capacity to Within One Bit},'' \emph{{IEEE} Trans. Inf. Theory}, vol.~54, pp.
  5534--5562, Dec. 2008.

\bibitem{Maddah-Ali:2008}
M.~A. Maddah-Ali, A.~S. Motahari, and A.~K. Khandani, ``{Communication over
  MIMO X channels: Interference alignment, decomposition, and performance
  analysis},'' \emph{{IEEE} Trans. Inf. Theory}, vol.~54, pp. 3457--3470, Aug.
  2008.

\bibitem{IA:conventional:2008}
V.~R. Cadambe and S.~A. Jafar, ``{Interference alignment and degrees of freedom
  of the K-user interference channel},'' \emph{{IEEE} Trans. Inf. Theory},
  vol.~54, pp. 3425--3441, Aug. 2008.

\bibitem{Ergodic:alignment:2009}
B.~Nazer, M.~Gastpar, S.~A. Jafar, and S.~Vishwanath, ``{Ergodic interference
  alignment},'' in \emph{Proc. ISIT}, June-July 2009.

\bibitem{IA:distributed:2008}
K.~S. Gomadam, V.~R. Cadambe, and S.~A. Jafar, ``{Approaching the capacity of
  wireless networks through distributed interference alignment},'' \emph{{To
  Appear in the IEEE Trans. Inf. Theory}}, 2011. Available:
  http://newport.eecs.uci.edu/~syed/papers/dist.pdf.

\bibitem{maximum:sum:rate}
I.~Santamaria, O.~Gonzalez, R.~W.~H. Jr., and S.~W. Peters, ``{Maximum sum-rate
  interference alignment algorithms for MIMO channels},'' in \emph{IEEE Proc.
  Globecom}, Dec. 2010.

\bibitem{alter:VT:2011}
S.~W. Peters and J.~R.~W.~Heath, ``{Cooperative algorithms for MIMO
  interference channels},'' \emph{{IEEE} Trans. Veh. Technol.}, vol.~60, pp.
  206--218, Jan. 2011.

\bibitem{IA:analysis}
{D. A. Schmidt and W. Utschick and M. L. Honig}, ``{Large system performance of
  interference alignment in single-beam MIMO networks},'' in \emph{IEEE Proc.
  Globecom}, Dec. 2010.

\bibitem{Feasibility:MIMO:2009}
C.~M. Yetis, T.~Gou, S.~A. Jafar, and A.~H. Kayran, ``{On feasibility of
  interference alignment in MIMO interference networks},'' \emph{{IEEE} Trans.
  Signal Process.}, vol.~58, pp. 4771--4782, Sep. 2010.

\bibitem{Many2one:2009}
G.~Bresler, A.~Parekh, and D.~N.~C. Tse, ``{The approximate capacity of the
  many-to-one and one-to-many gaussian interference channels},'' \emph{{IEEE}
  Trans. Inf. Theory}, vol.~56, pp. 4566--4592, Sep. 2010.

\bibitem{cadambe:lattice:2009}
V.~R. Cadambe, S.~A. Jafar, and S.~S. (Shitz), ``{Interference alignment on the
  deterministic channel and application to fully connected Gaussian
  interference networks},'' \emph{{IEEE} Trans. Inf. Theory}, vol.~55, pp.
  269--274, Jan. 2009.

\bibitem{Real:2009}
A.~S. Motahari, S.~O. Gharan, M.~Maddaha-Ali, and A.~K. Khandani, ``{Real
  interference alignment: Exploiting the potential of single antenna
  systems},'' 2009. Available: http://arxiv.org/abs/0908.2282v2.

\bibitem{Real:MIMO:2009}
A.~Ghasemi, A.~S. Motahari, and A.~K. Khandani, ``{Interference alignment for
  the K user MIMO interference channel},'' in \emph{ISIT 2010, Austin, Texas,
  U.S.A.}, June 2010.

\bibitem{Symmetric:2008}
S.~Sridharan, A.~Jafarian, S.~Vishwanath, and S.~A. Jafar, ``{Capacity of
  symmetric K-user Gaussian very strong interference channels},'' in \emph{IEEE
  Proc. Globecom}, Dec. 2008.

\bibitem{Threeusers:2008}
S.~Sridharan, A.~Jafarian, S.~Vishwanath, S.~A. Jafar, and S.~Shamai, ``{A
  layered lattice coding scheme for a class of three user Gaussian interference
  channels},'' Proceedings of 46th Annual Allerton Conference on Communication,
  Control and Computing, Sep 2008.

\bibitem{sdr:2002}
W.~K. Ma, T.~N. Davidson, K.~M. Wong, Z.~Q. Luo, and P.~C. Ching,
  ``{Quasi-maximum-likelihood multiuser detection using semi-definite
  relaxation with application to synchronous CDMA},'' \emph{{IEEE} Trans.
  Signal Process.}, vol.~50, pp. 912--922, Apr. 2002.

\bibitem{sdr:2004}
W.~K. Ma, P.~C. Ching, and Z.~Ding, ``{Semidefinite relaxation based multiuser
  detection for M-ary PSK multiuser systems},'' \emph{{IEEE} Trans. Signal
  Process.}, vol.~52, pp. 2862--2872, Oct. 2004.

\bibitem{sdr:mimo:2005}
A.~Wiesel, Y.~C. Eldar, and S.~Shamai, ``{Semidefinite relaxation for detection
  of 16-QAM signaling in MIMO channels},'' \emph{{IEEE} Signal Process. Lett.},
  vol.~12, pp. 653--656, Sep. 2005.

\bibitem{sdr:mimo:2005_2}
J.~Jalden, B.~Ottersten, and W.~K. Ma, ``{Reducing the average complexity of ML
  detection using semidefinite relaxation},'' in \emph{IEEE Proc. ICASSP' 09},
  March 2005.

\bibitem{ITU:1997}
{Recommendation ITU-R M.1225}, ``{Guidelines for evaluation of radio
  transmission technologies for IMT-2000},'' 1997.

\bibitem{Tresch:2009}
R.~Tresch, M.~Guillaud, and E.~Riegler, ``{On the achievability of interference
  alignment in the K-user constant MIMO interference channel},'' in
  \emph{IEEE/SP 15th Workshop on Statistical Signal Processing}, Aug.-Sep.
  2009.

\bibitem{comb:1998}
C.~H. Papadimitriou and K.~Steiglitz, \emph{{Combinatorial Optimization :
  Algorithms and Complexity }}.\hskip 1em plus 0.5em minus 0.4em\relax Mineola,
  N.Y. : Dover Edition, 1998.

\bibitem{comb:1988}
G.~L. Nemhauser and L.~A. Wolsey, \emph{{Integer and Combinatorial
  Optimization}}.\hskip 1em plus 0.5em minus 0.4em\relax New York: Wiley, 1988.

\bibitem{Simon:2005}
M.~K. Simon and M.~S. Alouini, \emph{{Digital Communication over Fading
  Channels (2nd Edition)}}.\hskip 1em plus 0.5em minus 0.4em\relax Hoboken,
  N.J. : John Wiley \& Sons,, 2005.

\bibitem{proakis:2001}
J.~G. Proakis, \emph{{Digital Communications (Fourth Edition)}}.\hskip 1em plus
  0.5em minus 0.4em\relax Boston : McGraw-Hill, 2001.

\bibitem{Convex:2004}
S.~Boyd and L.~Vandenberghe, \emph{{Convex Optimization}}.\hskip 1em plus 0.5em
  minus 0.4em\relax Cambridge, U.K.: Cambridge University Press, 2004.

\end{thebibliography}

\newpage

\end{document}